\documentclass{amsart} 

\usepackage{url,hyperref,lineno,microtype,subcaption}
\usepackage{algorithmic}
\usepackage{lipsum}
\usepackage{amsfonts}
\usepackage{graphicx}
\usepackage{epstopdf}
\usepackage{sidecap}
\usepackage{amsopn}
\usepackage{fullpage}
\usepackage{hyperref}

\numberwithin{equation}{section}

\DisableLigatures{encoding = *, family = * }

\begin{document}
	
\title{A stochastic approach to the synchronization of coupled oscillators}\thanks{This project has received funding from the European Research Council (ERC) under the European Union’s Horizon 2020 research and innovation programme (grant agreement NO. 694126-DyCon). The work of both authors was partially supported by the Grant MTM2017-92996-C2-1-R COSNET of MINECO (Spain) and by the Air Force Office of Scientific Research (AFOSR) under Award NO. FA9550-18-1-0242. The work of E.Z. was partially funded by the Alexander von Humboldt-Professorship program, the European Unions Horizon 2020 research and innovation programme under the Marie Sklodowska-Curie grant agreement No.765579-ConFlex, the Grant ICON-ANR-16-ACHN-0014 of the French ANR and the Transregio 154 Project ``Mathematical Modelling, Simulation and Optimization Using the Example of Gas Networks'' of the German DFG} 

\author{Umberto Biccari}
\address{Umberto Biccari, Chair of Computational Mathematics, Fundaci\'on Deusto, Avda. de las Universidades 24, 48007 Bilbao, Basque Country, Spain.}
\address{Umberto Biccari, Universidad de Deusto, Avda Universidades 24, 48007 Bilbao, Basque Country, Spain.}
\email{umberto.biccari@deusto.es,u.biccari@gmail.com}
	
\author{Enrique Zuazua}
\address{Enrique Zuazua, Chair in Applied Analysis, Alexander von Humboldt-Professorship, Department of Mathematics Friedrich-Alexander-Universit\"at Erlangen-N\"urnberg, 91058 Erlangen, Germany.}
\address{Enrique Zuazua, Chair of Computational Mathematics, Fundaci\'on Deusto, Avda. de las Universidades 24, 48007 Bilbao, Basque Country, Spain.}
\address{Enrique Zuazua, Departamento de Matem\'aticas, Universidad Aut\'onoma de Madrid, 28049 Madrid, Spain.}
\email{enrique.zuazua@fau.de}
	
\begin{abstract}
This paper deals with an optimal control problem associated with the Kuramoto model describing the dynamical behavior of a network of coupled oscillators. Our aim is to design a suitable control function allowing us to steer the system to a synchronized configuration in which all the oscillators are aligned on the same phase. This control is computed via the minimization of a given cost functional associated with the dynamics considered. For this minimization, we propose a novel approach based on the combination of a standard Gradient Descent (GD) methodology with the recently-developed Random Batch Method (RBM) for the efficient numerical approximation of collective dynamics. Our simulations show that the employment of RBM improves the performances of the GD algorithm, reducing the computational complexity of the minimization process and allowing for a more efficient control calculation.
\end{abstract}	
	
	
\maketitle

\section{Introduction}

Synchronization is a common phenomenon which has been observed in biological, chemical, physical and social systems for centuries and has attracted the interest of researcher in a diversified spectrum of scientific fields.

Common examples of synchronization phenomena often cited in review articles include groups of synchronously 
chirping crickets (\cite{walker1969acoustic}), fireflies flashing in unison (\cite{buck1988synchronous}), superconducting Josephson junction  (\cite{wiesenfeld1998frequency}), or crowds of people walking together that will tend to synchronize their footsteps (\cite{strogatz2005crowd}). 

Roughly speaking, synchronization means that a network of several periodic processes with different natural frequencies reaches an equilibrium configuration sharing the same common frequency as a result of their mutual interaction. 

This concept is closely related to the one of consensus for multi-agent systems, widely analyzed in many different frameworks including collective behavior of flocks and swarms, opinion formation, and distributed computing (see \cite{ben2005opinion,biccari2019dynamics,mehyar2005distributed,olfati2006flocking,olfati2007consensus}). In broad terms, consensus means to reach an agreement regarding a certain quantity of interest that depends on the state of all agents.

Synchronization is also a key issue in power electrical engineering, for instance in the model and stability analysis of utility power grids (\cite{chassin2005evaluating,filatrella2008analysis,sachtjen2000disturbances,strogatz2001exploring}). Indeed, large networks of connected power plants need to be synchronized to the same frequency in order to work properly and prevent the occurrence of blackouts.

Synchronization phenomena are most often characterized by the so-called Kuramoto model (\cite{kuramoto1975self}), describing the dynamical behavior of a (large) network of oscillators in a \textit{all-to-all} coupled configuration in which every oscillator is connected with all the others. This model extends the original studies by Winfree in the context of mutual synchronization in multi-oscillator systems based on a phase description (\cite{winfree1967biological}). In particular, in Kuramoto's work, synchronization appears as an asymptotic pattern which is spontaneously reached by the system when the interactions among the oscillators are sufficiently strong.

In some more recent contribution, control theoretic methods have been employed to analyze the synchronization phenomenon. For instance, in \cite{chopra2006passivity} the authors design passivity-based controls for the synchronization of multi-agent systems, with application to the general problem of multi-robot coordination. In \cite{sepulchre2005collective}, feedback control laws for the stabilization of coupled oscillators are designed and analyzed. In \cite{rosenblum2004delayed,tukhlina2007feedback}, the authors propose methods for the suppression of synchrony in a globally coupled oscillator network, based on (possibly time-delayed) feedback schemes. Finally, \cite{nabi2011single} deals with the problem of desynchronizing a network of synchronized and globally coupled neurons using an input to a single neuron. This is done in the spirit of dynamic programming, by minimizing a certain cost function over the whole state space.

In this work, we address the synchronization problem for coupled oscillators through the construction of a suitable control function via an appropriate optimization process. To this end, we propose a novel approach which combines a 
standard Gradient Descent (GD) methodology with the recently-developed Random Batch Method (RBM, see \cite{jin2020random}) for the efficient numerical approximation of collective dynamics. This methodology has the main advantages of allowing to significantly reduce the computational complexity of the optimization process, especially when considering oscillator networks of large size, yielding to an efficient control calculation.

At this regard, we shall mention that GD methodologies have already been applied in the context of the Kuramoto model. For instance, in \cite{taylor2016synchronization}, the author develop GD algorithms to efficiently solve optimization problems that aim to maximize phase synchronization via network modifications. Moreover, in \cite{markdahl2020high}, optimization and control theory techniques are applied to investigate the synchronization properties of a generalized Kuramoto model in which each oscillator lives on a compact, real Stiefel manifold. Nevertheless, to the best of our knowledge, the employment of stochastic techniques such as RBM to improve the efficiency of the GD strategy has never been proposed in the context of the Kuramoto model.

For completeness, let us stress that stochastic approaches have been widely considered, especially by the machine learning community, for treating minimization problems depending on very large data set. In this context, they have shown amazing performances in terms of the computational efficiency (see, for instance, \cite{bottou2018optimization} and the references therein). Nowadays, stochastic techniques are among the preeminent optimization methods in fields like empirical risk minimization (\cite{shalev2014accelerated}), data mining (\cite{toscher2010collaborative}) or artificial neural networks (\cite{schmidhuber2015deep}).

This contribution is organized as follows: in Section \ref{math_sec}, we present the Kuramoto model and we discuss some of its more relevant properties. We also provide there a rigorous mathematical characterization of the synchronization phenomenon. In Section \ref{control_sec}, we introduce the controlled Kuramoto model and we describe the GD methodology for the control computation. Moreover, we briefly present the RBM approach and its inclusion into the GD algorithm. Section \ref{numerics_sec} is devoted to the numerical simulations and to the comparison of the two optimization techniques considered in this paper. Finally, in Section \ref{conclusions_sec} we summarize and discuss our results.

\section{The mathematical model}\label{math_sec}

From a mathematical viewpoint, synchronization phenomena are most often described through the so-called Kuramoto model, consisting of a population of $N\geq 2$ coupled oscillators whose dynamics are governed by the following system of non-linear first-order ordinary differential equations 
\begin{align}\label{kuramoto_intro}
	\begin{cases}
		\displaystyle \dot{\theta}_i(t) = \omega_i + \frac{K}{N}\sum_{j=1}^N \sin \big(\theta_j(t)-\theta_i(t)\big),\quad i = 1,\ldots,N,\quad t>0
		\\
		\theta_i(0) = \theta_i^0,
	\end{cases}	
\end{align}
where $\theta_i(t)$, $i = 1,\ldots,N$, is the phase of the $i$-th oscillator, $\omega_i$ is its natural frequency and $K$ is the coupling strength.

The frequencies $\omega_i$ are assumed to be distributed with a given probability density $f(\omega)$, unimodal and symmetric around the mean frequency 
\begin{align*}
	\Omega = \frac 1N \sum_{i=1}^N \omega_i,
\end{align*}
that is, $f(\Omega+\omega) = f(\Omega-\omega)$.

In this framework, each oscillator tries to run independently at its own frequency, while the coupling tends to synchronize it to all the others.

In the literature, many notions of synchronization have been considered. For identical oscillators (i.e., those in which $\omega_i = \widehat{\omega}$ for every $i=1,\ldots,N$), one often studies whether the network can reach a configuration in which all the phases converge to the same value, that is 
\begin{align}\label{synchronization2}
	\lim_{t\to +\infty} |\theta_i(t)-\theta_j(t)| = 0, \quad \textrm{ for all }\; i,j = 1,\ldots,N.
\end{align}

For systems with heterogeneous dynamics, such as when the natural frequencies $\omega_i$ are not all identical (which is typical in real-world scenarios), this definition of synchronization is too restrictive (see \cite{sun2009master}). In these cases, \eqref{synchronization2} is replaced by the alignment condition
\begin{align}\label{synchronization}
	\lim_{t\to +\infty} |\dot{\theta}_i(t)-\dot{\theta}_j(t)| = 0, \quad \textrm{ for all }\; i,j = 1,\ldots,N,
\end{align}
according to which synchronization occurs when the phase differences given by $|\theta_i(t)-\theta_j(t)|$ become constant asymptotically for all $i,j\in1,\ldots,N$. This notion \eqref{synchronization}, which in some references is called \textit{complete synchronization} (see for instance \cite{ha2016emergence}), is the one that we will consider in this work.

In its original work \cite{kuramoto1975self}, Kuramoto considered the continuum limit case where $N\to +\infty$ and showed that the coupling $K$ has a key role in determining whether a network of oscillators can synchronize. In more detail, he showed that, when the coupling $K$ is weak, the oscillators run incoherently, whereas beyond a certain threshold collective synchronization emerges. 

Later on several research works provided specific bounds for the threshold of $K$ ensuring synchronization (see, e.g., \cite{acebron2005kuramoto,chopra2005synchronization,chopra2009exponential,dorfler2010synchronization,dorfler2013synchronization,jadbabaie2004stability}). In particular, in order to achieve \eqref{synchronization} it is enough that
\begin{align}\label{K_est}
	K > K^\ast =|\omega_{max}-\omega_{min}|,
\end{align} 
where $\omega_{min}<\omega_{max}$ are the minimum and maximum natural frequencies. 

Notice, however, that \eqref{synchronization} is an asymptotic characterization, meaning that is satisfied as $t\to +\infty$. In this work we are rather interested in the possibility of synchronizing the oscillators in a finite time horizon $T$. As we will discuss in the next section, this may be achieved by introducing a control into the Kuramoto model \eqref{kuramoto_intro}. 

\section{Optimal control of the Kuramoto model}\label{control_sec}

As we mentioned in Section \ref{math_sec}, in this work we are interested in the finite-time synchronization of the Kuramoto model. In particular, we aim at designing a control capable to steer the Kuramoto dynamics \eqref{kuramoto_intro} to synchronization in a final time horizon $T$. In other words, we are going to consider the controlled system
\begin{align}\label{kuramoto_control}
	\begin{cases}
		\displaystyle\dot{\theta}_i(t) = \omega_i + \frac{Ku(t)}{N}\sum_{j=1}^N \sin\big(\theta_j(t)-\theta_i(t)\big),\quad i = 1,\ldots,N,\quad t>0
		\\
		\theta_i(0) = \theta^0_i,
	\end{cases}
\end{align}
and we want to compute a control function $u$ such that the synchronized configuration \eqref{synchronization} is achieved at time $T$, i.e., 
\begin{align}\label{consensus}
	|\dot{\theta}_i(T)-\dot{\theta}_j(T)| = 0, \quad \textrm{ for all } i,j = 1,\ldots,N.
\end{align}

From the practical applications viewpoint, this problem may be assimilated for instance to the necessity of synchronizing all the components of an electric grid after a black-out. In this interpretation, the different elements in the grid are represented by the oscillators in \eqref{kuramoto_control}, and $T$ is the time horizon we provide for the black-start, being therefore an external input to our problem. The objective is then to complete restoring the network in a finite (possibly small) time $T$, which can be done by introducing a control in the system.

To compute this optimal control allowing us to reach the synchronized configuration \eqref{consensus} we will adopt a classical optimization approach based on the resolution of the following optimization problem 
\begin{align}\label{functional}
	&\widehat{u} = \min_{u\in L^2(0,T;\mathbb{R})} J(u)\notag 
	\\
	&J(u) = \frac{1}{2} \sum_{i,j=1}^N \sin^2\big(\theta_j(T)-\theta_i(T)\big) + \frac \beta2 \|u\|^2_{L^2(0,T;\mathbb{R})},
\end{align}
subject to the dynamics \eqref{kuramoto_control}. Here, with $L^2(0,T;\mathbb{R})$ we denoted the space of all functions $u:(0,T)\to\mathbb{R}$ for which the following norm is finite:
\begin{align}\label{L2norm}
	\|u\|_{L^2(0,T;\mathbb{R})} := \left(\int_0^T |u(t)|^2\,dt\right)^{\frac 12}.
\end{align}
In what follows, we will use the abridged notation $\|u\|_2 := \|u\|_{L^2(0,T;\mathbb{R})}$.

In the cost functional \eqref{functional}, the first term enhances the fact that all the oscillators have to synchronize at time $T$. In particular, the optimal control $\widehat{u}$ will yield to a dynamics in which
\begin{align}\label{consensus_sinus}
	\sin(\theta_j(T)-\theta_i(T)) = 0 \quad \Rightarrow \quad \theta_j(T)-\theta_i(T) = k\pi, \; k\in\mathbb{Z}.
\end{align}

This is consistent with \eqref{consensus}. For completeness, let us also stress that, in the case of identical oscillators, it has been shown for instance in \cite{ha2015remarks} that, at least asymptotically, the two notions \eqref{consensus} and \eqref{consensus_sinus} coincide.

The second term in \eqref{functional} is introduced to avoid controls with a too large size. In it, $\beta>0$ is a (usually small) penalization parameter which allows to tune the norm of the optimal control $\widehat{u}$. Roughly speaking, the smaller is $\beta$ the larger will be $\widehat{u}$. A more detailed discussion on this point will be presented in Section \ref{numerics_sec}.

Through the minimization of $J(u)$, we will obtain a unique scalar control function $\widehat{u}:(0,T)\to \mathbb{R}$, $\widehat{u}>1$, for all the oscillators included in the network. In other words, we are going to define a unique control law which is capable to act globally on the entire oscillator network in order to reach the desired synchronized configuration. This is a different approach than the ones presented in \cite{chopra2006passivity,nabi2011single,rosenblum2004delayed,sepulchre2005collective,tukhlina2007feedback} which we mentioned above and are based on designing feedback laws or controlling only some specific components of the model, using the coupling to deal with the uncontrolled ones.

One advantage of the control strategy that we propose is that, requiring only one control computation, from the computational viewpoint is more efficient than a feedback approach which necessitates repeated measurements of the state. Moreover, let us notice that, in \eqref{kuramoto_control}, the control acts as a multiplicative force which increases the coupling among the oscillators, thus enhancing their synchronization properties. In particular, as we will see in our numerical simulations, this will allow to reach synchronization also in situations where $K$ violates the condition \eqref{K_est} and the uncontrolled dynamics runs incoherently towards a desynchronized configuration.

Nevertheless, our proposed methodology may have the disadvantage of being less flexible than the others we mentioned above. In particular, it does not allow to control only a specific component of the network and this may be a limitation in certain practical applications. 

Let us stress that the above considerations are merely heuristic and should be corroborated by a deeper analysis based, for instance, on sharp numerical experiments. Notwithstanding that, in the present work we will not address this specific issue, since our main interest is not to compare the performances of different control strategies but rather to present an efficient way to tackle the control problem \eqref{kuramoto_control}.
  
In the optimization literature, several different techniques have been proposed for minimizing the functional $J(u)$ (see, e.g., \cite{nocedal2006numerical}). In this work, we focus on the standard GD method, which looks for the minimum $u$ as the limit $k\to +\infty$ of the following iterative process
\begin{align}\label{GD_scheme}
	u^{k+1} = u^k - \eta_k\nabla J(u^k),
\end{align}
where $\eta_k>0$ is called the step-size or, in the machine learning context, the learning rate. The step size is typically selected to be a constant depending on certain key parameters of the optimization problem, or following an adaptive strategy. See, e.g., \cite[Section 1.2.3]{nesterov2004applied} for more details.

This gradient technique is most often chosen because it is easy to implement and not very memory demanding. 

Nevertheless, when applied to the optimal control of collective dynamics, the GD methodology has a main drawback. Indeed, as we shall see in  Section \ref{GD_sec}, at each iteration $k$ the optimization scheme \eqref{GD_scheme} requires to solve \eqref{kuramoto_control}, that is, a $N$-dimensional non-linear dynamical system. This may rapidly become computationally very expensive, especially when the number $N$ of oscillators in our system is large. 

In order to reduce this computational burden, in this work we propose a novel methodology which combines the standard GD algorithm with the so-called \textit{Random Batch Method} (RBM). 

RBM is a recently developed approach which has been introduced in \cite{jin2020random} for the numerical simulation of high-dimensional collective behavior problems. This method uses small but random batches for particle interactions, lowering the computational cost $\mathcal O(N^2)$ per time step to $\mathcal O (N)$, for systems with $N$ particles with binary interactions. Therefore, as our numerical simulations will confirm, embedding RBM into the GD iterative scheme yields to a less expensive algorithm and, consequently, to a more efficient control computation. In what follows, we will call this approach GD-RBM algorithm, to differentiate it from the standard GD one.	

\subsection{The Gradient Descent approach}\label{GD_sec}

Let us now describe in detail the GD approach to minimize the functional \eqref{functional}, and discuss its convergence properties. 

In order to fully define the iterative scheme \eqref{GD_scheme}, we need to compute the gradient $\nabla J(u)$. Since we are dealing with a non-linear control problem, we will do this via the so-called \textit{Pontryagin maximum principle} (see \cite[Chapter 4, Section 4.8]{troltzsch2010optimal} or \cite[Chapter 7]{trelat2005controle}).

To this end, let us first rewrite the dynamics \eqref{kuramoto_control} in a vectorial form as follows
\begin{align}\label{kuramoto_vec}
	\begin{cases}
		\dot{\Theta}(t) = \Omega + F\big(\Theta(t),u(t)\big), \quad t>0
		\\
		\Theta(0) = \Theta^0,
	\end{cases} 
\end{align}
with $\Theta :=(\theta_1,\ldots,\theta_N)^\top$, $\Theta^0:=(\theta_1^0,\ldots,\theta_N^0)^\top$ and $\Omega :=(\omega_1,\ldots,\omega_N)^\top$, and where $F$ is the vector field given by
\begin{align}\label{vectorField}
	F = (F_1,\ldots,F_N), \quad F_i:= \frac{Ku(t)}{N}\sum_{j=1}^N\sin\big(\theta_j(t)-\theta_i(t)\big), \quad i=1,\ldots,N.
\end{align}
In the control literature, \eqref{kuramoto_vec} is usually called the \textit{primal system}.

Using the notation just introduced, we can then see that $J(u)$ can be rewritten in the form
\begin{align}\label{functionalP}
	J(u) = \int_0^T L(u(t))\,dt + \phi(\Theta(T)),
\end{align}
with
\begin{align*}
	L(u(t)) = \frac \beta2 |u(t)|^2 \quad \textrm{ and } \quad \phi(\Theta(T)) = \frac 12\sum_{j=1}^N \sin^2\big(\theta_j(T)-\theta_i(T)\big).
\end{align*}

Let us stress that \eqref{functionalP} is in the standard form to apply the Pontryagin maximum principle. Through this approach, we can obtain the following expression for the gradient of $J(u)$
\begin{align}\label{gradJ}
	\nabla J(u) = \beta u + (\mathcal D_uF)^\top p,
\end{align}
where $\mathcal D_uF$ indicates the Jacobian of the vector field $F$, computed with respect to the variable $u$. 

In \eqref{gradJ}, we denoted with $p = (p_1,\ldots,p_N)$ the solution of the \textit{adjoint equation} associated with \eqref{kuramoto_control}, which is given by
\begin{align}\label{adjoint_compact}
	\begin{cases}
		-\dot{p} = (\mathcal D_{\Theta}F)^\top p
		\\
		p(T) = \nabla_{\Theta(T)}\phi(\Theta(T)),
	\end{cases}
\end{align}
where $\mathcal D_\Theta F$ stands again for the Jacobian of the vector field $F$, this time computed with respect to the variable $\Theta$. 

Taking into account the expression \eqref{vectorField} of the vector field $F$, we can then readily check that the iterative scheme \eqref{GD_scheme} becomes
\begin{align}\label{GD_scheme_expl}
	u^{k+1} = u^k -\eta_k\left[\beta u^k + \frac KN\sum_{i=1}^N p_i\left(\sum_{j=1}^N \sin(\theta_j-\theta_i)\right)\right],
\end{align}
with 
\begin{align}\label{adjoint}
	\begin{cases}
		\displaystyle-\dot{p}_i = -\frac{Kup_i}{N}\sum_{i\neq j=1}^N \cos\big(\theta_j-\theta_i\big) + \frac{Ku}{N}\sum_{i\neq j=1}^N p_j \cos\big(\theta_j-\theta_i\big),\quad i = 1,\ldots,N,\quad t>0
		\\[20pt]
		\displaystyle p_i(T) = \frac 12\sum_{i\neq j=1}^N \sin\big(2\theta_i(T)-2\theta_j(T)\big).
\end{cases}
\end{align}

In view of the above computations, the GD algorithm for the minimization of the cost functional $J(u)$ can be explicitly formulated as follows:

\paragraph*{GD algorithm}

\begin{algorithmic}
	\STATE{\textbf{input} $\Theta^0$: initial condition of the primal system \eqref{kuramoto_vec}\\
			\quad \qquad $u^0$: initial guess for the control $u$ \\
			\quad \qquad $k\leftarrow 0$: iteration counter \\
			\quad \qquad $k_{max}$: maximum number of iterations allowed\\
			\quad \qquad $tol$: tolerance}
	\WHILE{STOP-CRIT and $k<k_{max}$}
		\STATE{$k\leftarrow k+1$}
	\FOR{$j=1$ to $N$}
		\STATE{Solve the the primal system \eqref{kuramoto_vec}}
		\STATE{Solve the the adjoint system \eqref{adjoint}}
	\ENDFOR
		\STATE{Update the control through the scheme \eqref{GD_scheme_expl}}
	\ENDWHILE
	\RETURN $u^{k+1} = \widehat{u}$: minimum of the functional $J(u)$.
\end{algorithmic}

In particular, we see that the control computation through the above algorithm requires, at each iteration $k$, to solve $2N$ non-linear differential equations ($N$ for the variables $\theta_i$ and $N$ for $p_i$). If we introduce the time-mesh of $N_t$ points
\begin{align*}
	0 = t_0<t_1<\ldots<t_{N_t} = T, \quad t_m = t_0 + m\frac{T}{N_t}, \;\; m = 1,\ldots,N_t,
\end{align*}
at each time-step $t_m$ this operation has a computational cost of $\mathcal O(N^2)$ and the total computational complexity  for the simulation of \eqref{kuramoto_control} and \eqref{adjoint} will then be $\mathcal O(N_tN^2)$. If $N$ is large, that is, if the number of oscillators in the network is considerable, this will rapidly become very expensive.

\subsection{The Random Batch Method}\label{RBM_sec}

In order to reduce the computational burden of GD for the optimization process \eqref{functional}, we propose a modification of this algorithm which includes the aforementioned \textit{Random Batch Method} (RBM) for the numerical simulation of the ODE systems \eqref{kuramoto_control} and \eqref{adjoint}.

This technique, presented in \cite{jin2020random} for interacting particle systems, is based on the following simple idea: at each time step $t_m = m\cdot dt$ in the mesh we employ to solve the dynamics, we divide randomly the $N$ particles into $n$ small batches with size $2\leq P<N$, denoted by $C_q$, $q = 1,\ldots,n$, that is
\begin{align*}
	& C_q = \{i_{q_1},\ldots,i_{q_P}\}\subset \{1,\ldots,N\}, & &\mbox{ for all } q = 1,\ldots,n
	\\
	& C_q\cap C_r = \emptyset, & & \mbox{ for all } q,r = 1,\ldots,n
	\\
	& \bigcup_{q = 1}^n C_q = \{1,\ldots,N\}.
\end{align*}
Notice that the last batch $C_n$ may have size smaller than $P$ if $nP\neq N$.

Once this partition of $\{1,\ldots,N\}$ has been performed, we solve the dynamics by interacting only particles within the same batch. This gives the following algorithm for the numerical  approximation of \eqref{kuramoto_control} and \eqref{adjoint}:

\paragraph*{RBM algorithm}

\begin{algorithmic}
	\FOR{$m=1$ to $N_t=T/dt$}
	\STATE{Divide randomly $\{1,\ldots,N\}$ into $n$ batches $C_q$, $q = 1,\ldots,n$}
	\FOR{$q=1$ to $n$}
	\STATE{Update $\theta_i$ $(i\in C_q)$ by solving the ODE
	\begin{align*}
		\begin{cases}
			\displaystyle\dot{\theta}_i = \omega_i + \frac{Ku}{P}\sum_{j\in C_q} \sin\big(\theta_j-\theta_i\big)
			\\
			\theta_i(0) = \theta^0_i.
		\end{cases}
	\end{align*}
	}
	\STATE{Update $p_i$ $(i\in C_q)$ by solving the ODE
	\begin{align*}
		\begin{cases}
			\displaystyle -\dot{p}_i = -\frac{Kup_i}{P}\sum_{j\in C_q} \cos\big(\theta_j-\theta_i\big) + \frac{Ku}{P}\sum_{i\neq j\in C_q} p_j \cos\big(\theta_j-\theta_i\big)
			\\
			\displaystyle p_i(T) = \frac 12\sum_{i\neq j\in C_q} \sin\big(2\theta_i(T)-2\theta_j(T)\big).
		\end{cases}
	\end{align*}
	}
	\ENDFOR
	\ENDFOR
\end{algorithmic}

Regarding the complexity, note that random division into $n$ batches of can be implemented using random permutation. In Matlab, this can be done by using the function \textit{randperm(N)}. Then, the first $P$ elements are considered to be in the first batch, the second $P$ elements are in the second batch, and so on. According to the discussion presented in \cite{jin2020random}, at each time step $t_m$ this procedure yields to a cost of $\mathcal O(PN)$ for approximating the dynamics with RBM.

If one is to simulate up to time $T$, the total number of time steps is $N_t$ as in the algorithm above. Then, the total computational complexity for the simulation of \eqref{kuramoto_control} and \eqref{adjoint} is $\mathcal O(PN_tN)$. Notice that, since $P<N$, this is always smaller than $\mathcal O(N_tN^2)$.

Summarizing, with the GD and GD-RBM methodologies we obtain the following per-iteration costs:
\begin{itemize}
	\item GD $\longrightarrow\mbox{cost}_{GD} = \mathcal C_{GD}N_tN^2$.
	\item GD-RBM $\longrightarrow\mbox{cost}_{GD-RBM} = \mathcal C_{GD-RBM} PN_tN$.
\end{itemize}

Therefore, independently of the value of $N$, employing RBM to simulate the dynamics in each iteration of GD yields improvements in terms of the computational cost.

For completeness, we shall mention that the above considerations are simply heuristic and would require a deeper analysis. As a matter of fact, to have a rigorous validation of the reduction in the computational complexity when using RBM one should have more precise information on the two constants $\mathcal C_{GD}$ and $\mathcal C_{GD-RBM}$, and be sure that the difference among them does not overwhelm the help that the batching procedure is providing. 

At this regard, let us stress that the RBM method has been developed only recently in \cite{jin2020random} and, at present time, there is not a well-established qualitative analysis on its computational cost, going in more detail than what we mentioned above. The evidence that in our case of the Kuramoto model \eqref{kuramoto_control} the GD-RBM method allows for a more efficient control computation, in particular for large oscillator networks, will then be given through the numerical simulations in Section \ref{numerics_sec}.

\subsection{Convergence analysis}\label{convergence_sec}

To complete this section, let us briefly comment about the convergence properties of the GD methodology. It is nowadays classically known that the convergence rate of the GD algorithm is determined by the regularity of the objective function. In our case, since $J(u)$ is L-smooth, that is
\begin{align*}
	\|\nabla J(u)-\nabla J(v)\|_2 \leq L \|u-v\|_2,
\end{align*}
it can be proven that 
\begin{align}\label{gradConvergence}
	\|\nabla J(u^k)\|_2 \to 0 \quad \mbox{ as } k\to +\infty 
\end{align}
and
\begin{align}\label{GDconvergence}
	\|J(u^k)-J(\widehat{u})\|_2 = \mathcal O\left(\frac 1k\right),
\end{align}
where, we recall, $\widehat{u}$ denotes the minimum of $J(u)$ and the norm $\|\cdot\|_{L^2(0,T;\mathbb{R})}$ has been defined in \eqref{L2norm}. 

In particular, \eqref{GDconvergence} implies that for achieving $\varepsilon$-optimality, i.e. for obtaining $\|J(u^k)-J(\widehat{u})\|_2 <\varepsilon$, the GD algorithm requires $k=\mathcal O(\varepsilon^{-1})$ iterations.

Combining this with the per-iteration costs we gave at the end of Section \ref{RBM_sec}, we can thus obtain the following total computational costs
\begin{itemize}
	\item GD: $\mathcal O\left(\frac{N_tN^2}{\varepsilon}\right)$
	\item GD-RBM: $\mathcal O\left(\frac{PN_tN}{\varepsilon}\right)$,
\end{itemize}
and we can conclude that the GD-RBM approach will be more efficient than the standard GD one to solve our optimization problem. This is enhanced for large values of $N$ and will be confirmed by our numerical simulations.

\section{Numerical simulations}\label{numerics_sec}

We present here our numerical results for the control of $N$ coupled oscillators described by the Kuramoto model \eqref{kuramoto_control}, following the strategy previously described. This section is divided into two parts:
\begin{itemize}
	\item[1.] In a first moment, we will show that the optimization problem \eqref{functional} indeed allows to compute an effective control function which is capable to steer the Kuramoto model \eqref{kuramoto_control} to a synchronized configuration. This will be done both for a strong coupling $K>K^\ast$ (see \eqref{K_est}) and for a weak coupling $K<K^\ast$. Besides, we will also briefly analyze the role of the parameter $\beta$ in the optimization process. Finally, we will show the efficacy of our control strategy in the more realistic cases of a sparse interaction network and for a second-order Kuramoto model with damping.
	\item[2.] Once the effectiveness of the control strategy we propose has been corroborated, we will compare the GD and GD-RBM algorithms for the minimization of $J(u)$. In particular, we will show how the RBM approach allows to significantly reduce the computational complexity of the GD algorithm for the calculation of the control $u$, especially when considering oscillator networks of large dimension.
\end{itemize}

The oscillators are chosen such that their natural frequencies are given following the normal probability law
\begin{align}\label{distribution}
	f(\omega) = \frac{1}{5\sigma\sqrt{2\pi}}e^{-\frac{\omega^2}{2\sigma}}, 
\end{align}
with $\sigma = 0.1$. This means that the values of $\omega_{min}$ and $\omega_{max}$ are given respectively by 
\begin{align*}
	&\omega_{\min} = \min_{\omega\in\mathbb{R}}f(\omega) = 0
	\\
	&\omega_{\max} = \max_{\omega\in\mathbb{R}}f(\omega) = f(0) = \frac{2}{\sqrt{2\pi}}
\end{align*}
and the coupling gain $K$ which is necessary for synchronization in the absence of a control has to satisfy (see \eqref{K_est})
\begin{align*}
	K > |\omega_{max}-\omega_{min}| = \frac{2}{\sqrt{2\pi}}.
\end{align*}

The initial datum $\theta^0$ is chosen following a normal distribution as well, in such a way that $|\theta^0_i-\theta^0_j|<2\pi$ for all $i,j=1,\ldots,N$. Let us stress that this choice of $\theta^0$ allows the synchronization of the uncontrolled model, as it has been shown for instance in \cite{dong2013synchronization}. 

Without loss of generality, we considered the time horizon $T=3s$ for completing the synchronization. That is, we want all the oscillators in our model to reach the configuration \eqref{consensus} in three seconds. 

Finally, we used an explicit Euler scheme for solving the direct and adjoint dynamics \eqref{kuramoto_control} and \eqref{adjoint} during the minimization of $J(u)$, and we chose as a stopping criterion 
\begin{align}\label{stop_crit}
	e_k:=\frac{\|\nabla J(u^{k})\|_{2}}{\|u^k\|_{2}} < \varepsilon,
\end{align}
with $\varepsilon = 10^{-4}$, and where the notation $\|\cdot\|_2$ stands again for the $L^2(0,T\;\mathbb{R})$-norm defined in \eqref{L2norm}. Let us stress that the stopping criterion \eqref{stop_crit} is consistent with \eqref{gradConvergence} and \eqref{GDconvergence}.

\subsection{Computation of the optimal control}

In this section, we show that through the optimization problem \eqref{functional} we are able to compute an effective control function which is capable to steer the Kuramoto model \eqref{kuramoto_control} to a synchronized configuration in a given time horizon $T$. 

We performed the simulations in Matlab R2018a on a laptop with Intel Core $i5-7200U CPU @ 2.50GHz\times4$ processor and $7.7$ GiB RAM.

We start by considering a simple scenario of $N=10$ oscillators in an all-to-all coupled configuration and with a coupling gain $K>K^\ast$. 
Moreover, we set the penalization parameter $\beta$ in \eqref{functional} to take the value $\beta=10^{-7}$

When using the GD-RBM approach, the family of $N=10$ oscillators has been separated into $n=5$ batches of size $P=2$.

\begin{figure}[h!]
	\centering
	\includegraphics[scale=1]{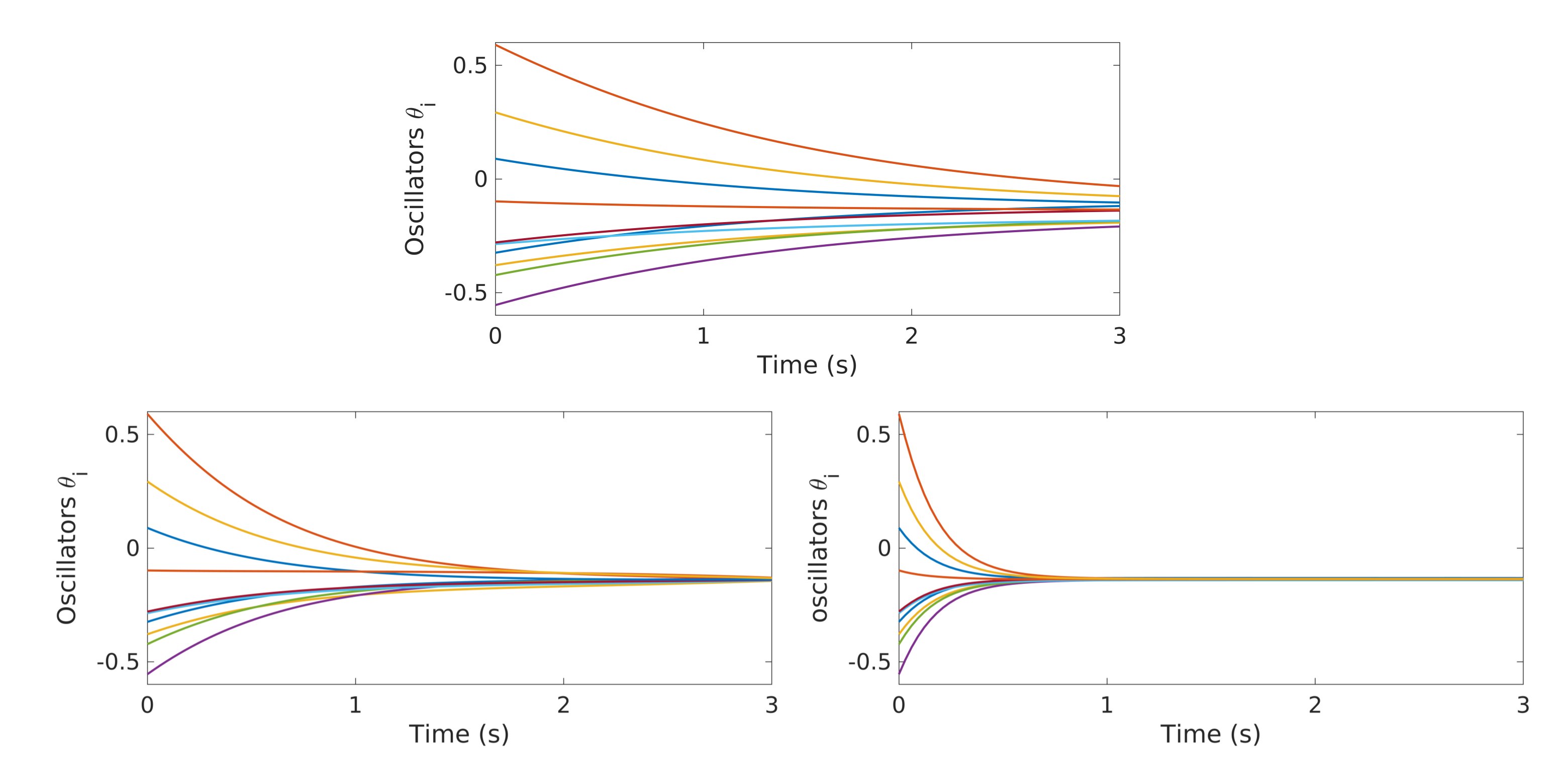}
	\caption{Top: evolution of the free dynamics of the Kuramoto model \eqref{kuramoto_control} with $N=10$ oscillators. Bottom: evolution of the controlled dynamics of the Kuramoto model \eqref{kuramoto_control} with $N=10$ oscillators. The control function $\widehat{u}$ is obtained with the GD (left) and the GD-RBM (right) approach.}\label{fig:1}
\end{figure}

In Figure \ref{fig:1}-top, we show the evolution of the uncontrolled dynamics, which corresponds to taking $u\equiv 1$ in \eqref{kuramoto_control}.
As we can see, the oscillators are evolving towards a synchronized configuration, which is consistent with our choice of the coupling gain $K$. Nevertheless, synchronization is not reached in the short time horizon we are providing. At this regard, let us remark that, for the uncontrolled Kuramoto model with a sufficiently strong coupling gain, synchronization is expected to be reached only asymptotically, i.e. when $t\to +\infty$. 

In Figure \ref{fig:1}-bottom, we show the evolution of the same dynamics, this time under the action of the control function $u$ computed through the minimization of $J(u)$. The subplot on the left corresponds to the simulations done with the GD approach, while the one on the right is done employing GD-RBM. We can clearly see how, in both cases, the oscillators are all synchronized at the final time $T=3s$. This means that both algorithms managed to compute an effective control.

\begin{figure}[h!]
	\centering 
	\includegraphics[scale=1]{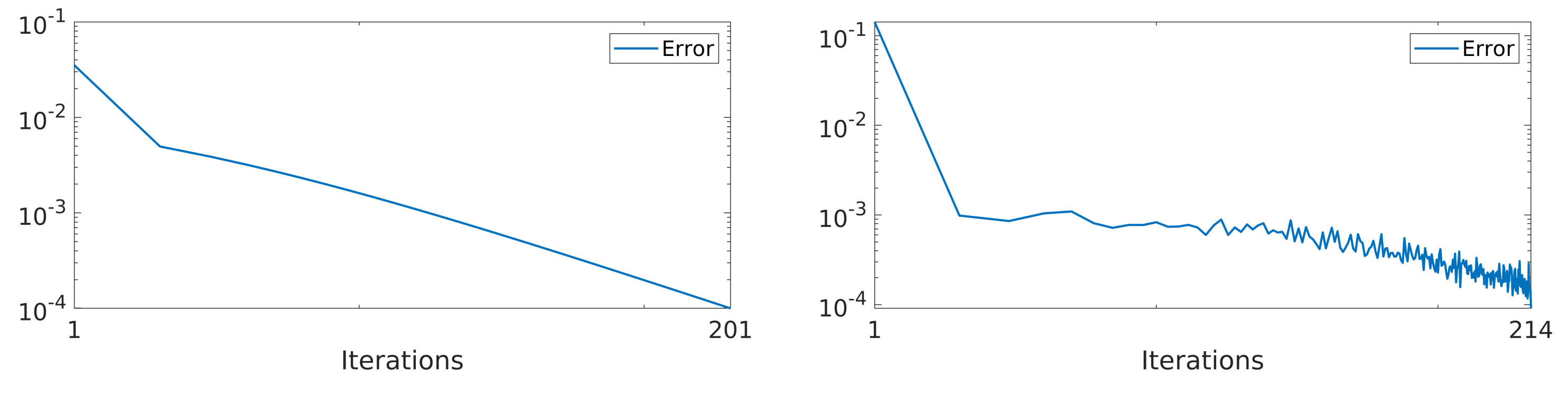}
	\caption{Convergence of the error $e_k$ (see \eqref{stop_crit}) in logarithmic scale with the GD (left) and GD-RBM (right) algorithm.}\label{fig:3}
\end{figure}

In Figure \ref{fig:3}, we show the convergence of the error in logarithmic scale when applying both the GD and GD-RBM approach. We can appreciate how, in the case of GD-RBM, this convergence is not monotonic as it is for the GD algorithm. This, however, is not surprising due to the stochastic nature of the RBM methodology.

\begin{figure}[!h]
	\centering
	\includegraphics[scale=0.45]{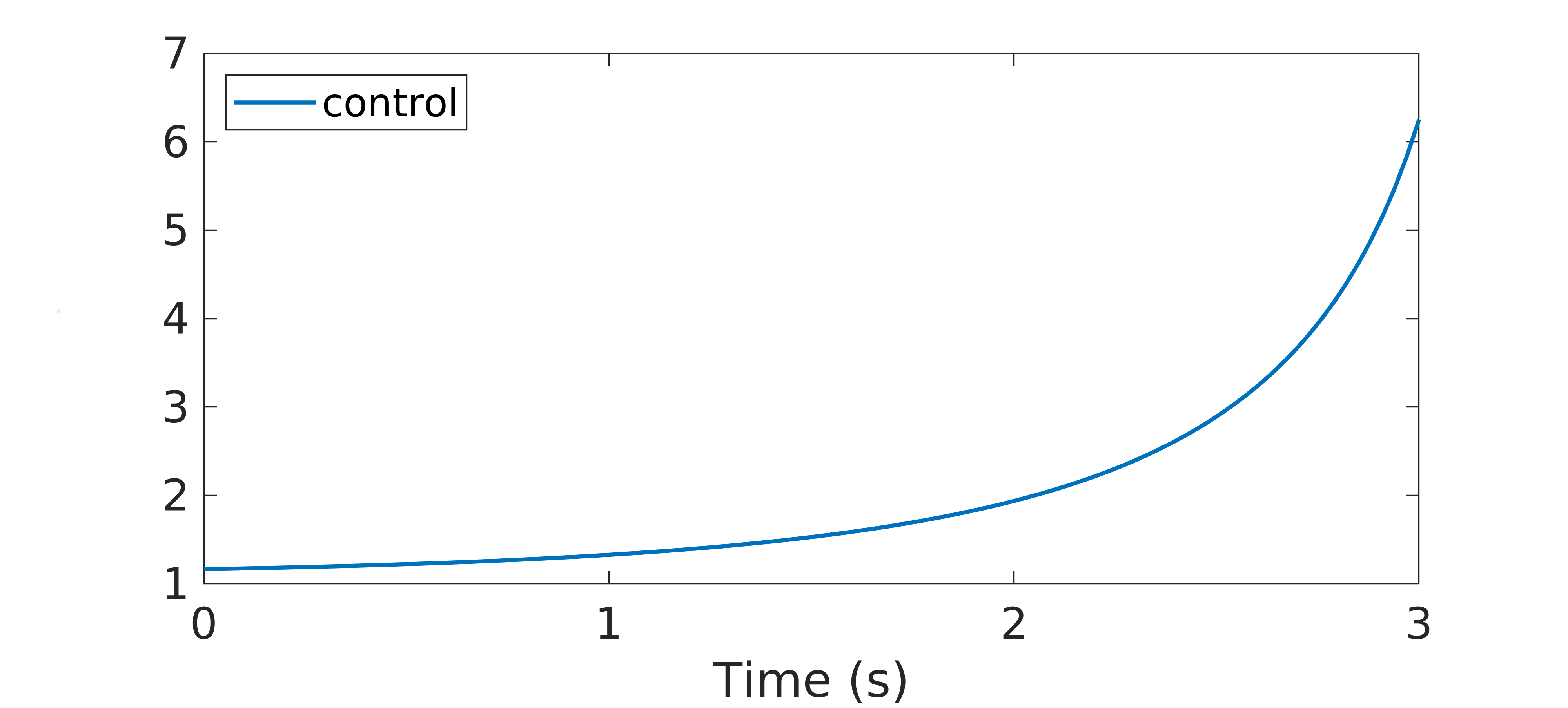}
	\caption{Control function $\widehat{u}$ obtained through the GD-RBM algorithm applied to the Kuramoto model \eqref{kuramoto_control} with $N=10$ oscillators.}\label{fig:4} 
\end{figure}

In Figure \ref{fig:4}, we display the behavior of the control function $\widehat{u}$ computed via the GD-RBM algorithm. We can see how, at the beginning of the time interval we are considering, this control is close to one and it is increasing with a small slope. On the other hand, this growth becomes more pronounced as we get closer to the final time $T=3s$.

Notice that, in \eqref{kuramoto_control}, $\widehat{u}$ enters as a multiplicative control which modifies the strength of the coupling $K$. Hence, according to the profile displayed in Figure \ref{fig:3}, the control function $\widehat{u}$ we computed is initially letting the system evolving following its natural dynamics. Then, as the time evolves towards the horizon $T=3s$, $\widehat{u}$ enhances the coupling strength $K$ in order to reach the desired synchronized configuration \eqref{consensus}.

Finally, notice also that the control $\widehat{u}$ is always positive. This is actually not surprising, if one takes into account the following observation. 

In the Kuramoto model, in order to reach synchronization the coupling strength $K$ needs to be positive. Otherwise, the system would converge to a desynchronized configuration (see \cite{hong2011kuramoto}). Moreover, according to the model \eqref{kuramoto_control}, if we start from $K>0$, in order to maintain this coupling positive $\widehat{u}$ has to remain positive as well.

Recall that $\widehat{u}$ is computed minimizing the functional \eqref{functional}, in which the second term is a measurement of the level of synchronization in the model. Hence, since negative values of $\widehat{u}$ would lead to desynchronization and to the corresponding increasing of the functional, these values remain automatically excluded during the minimization process.

Let us now discuss briefly the role of the penalization parameter $\beta$ in the computation of the optimal control. To this end, we have run simulations with different values of $\beta=10^{-2},10^{-3},10^{-4}$ and $10^{-7}$. 

As we already mentioned in Section \ref{control_sec}, in the cost functional \eqref{functional} $\beta$ is a (usually small) penalization parameter which allows to tune the norm of the optimal control $\widehat{u}$, that is, the amount of energy that the control introduces into the system. Roughly speaking, the smaller is $\beta$ the larger will be $\widehat{u}$. 

\begin{figure}[!h]
	\centering
	\includegraphics[scale=0.45]{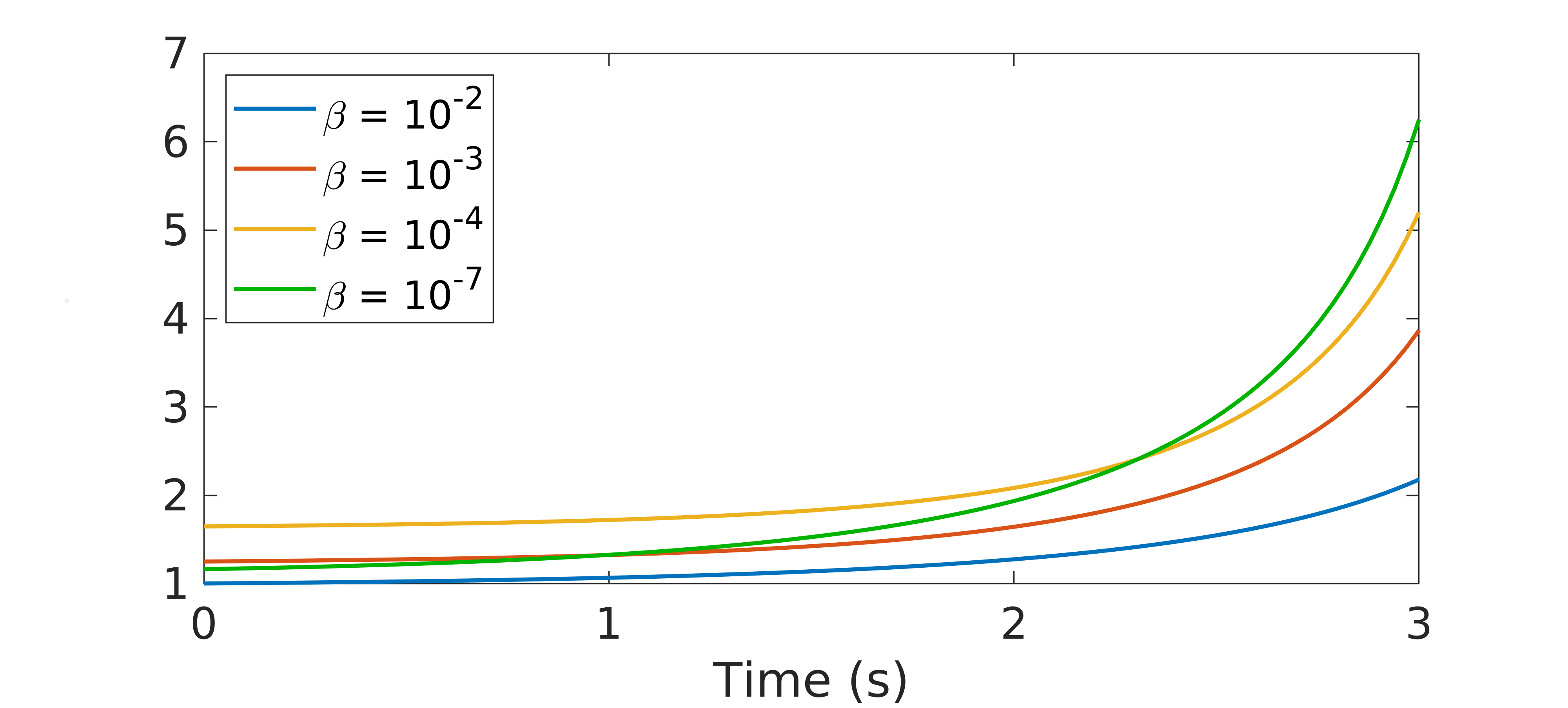}
	\caption{Control function $\widehat{u}$ obtained through the GD-RBM algorithm applied to the Kuramoto model \eqref{kuramoto_control} with $N=10$ oscillators and different values of $\beta$.}\label{fig:5} 
\end{figure}

This is clearly seen in Figure \ref{fig:5}. In particular, we can appreciate how, for $\beta=10^{-2}$, the computed control remains smaller than in the other cases. 

We already mentioned above that the effect of the control in \eqref{kuramoto_control} is to enhance synchronization by modifying the strength of the coupling $K$. Hence, we can expect that, if $\widehat{u}$ is small (in particular, if it remains close to one), the synchronization properties of the Kuramoto model \eqref{kuramoto_control} will be worst than when applying a larger control. At this regard, let us recall that the level of synchronization in \eqref{kuramoto_control} can be analyzed in terms of the quantity
\begin{align*}
	r(t):= \left|\frac 1N \sum_{j=1}^N e^{i\theta_j(t)}\right|,
\end{align*}
measuring the coherence of the oscillator population (see \cite{acebron2005kuramoto}). In particular we always have $0\leq r(t)\leq 1$ and synchronization arises when $r$ reaches the value one. 

\begin{figure}[h!]
	\centering 
	\includegraphics[scale=0.4]{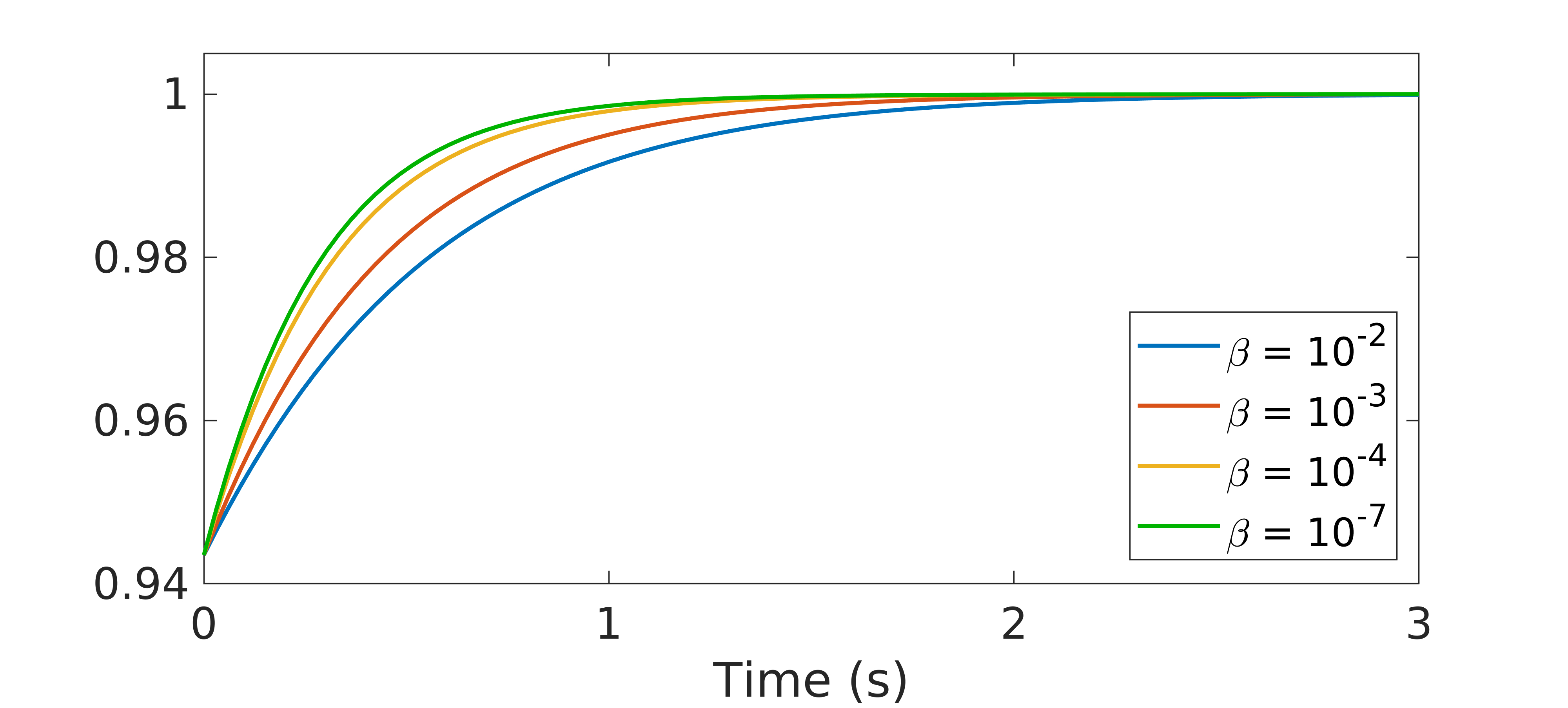}
	\caption{Behavior of the synchronization function $r(t)$ corresponding to the controlled dynamics \eqref{kuramoto_control} for different values of the parameter $\beta$.}\label{fig:6}
\end{figure}

In Figure \ref{fig:6}, we show the behavior of $r(t)$ with respect to the parameter $\beta$. On the one hand, in all the cases displayed we can clearly see that $r(T)=1$. This means that all the computed controls will be effective to steer the system \eqref{kuramoto_control} to its synchronized configuration at time $T$. On the other hand, we can also notice how, when decreasing $\beta$, the function $r(t)$ reaches the value one faster, meaning that the corresponding control is expected to yield to better synchronization properties.

Let us now conclude this section by showing that the control strategy that we propose in this paper is effective also in situations in which the coupling gain among the oscillators is too weak to ensure synchronization for the uncontrolled dynamics of the Kuramoto model. This corresponds to taking $K<K^\ast$ (see \eqref{K_est}). In particular, we will consider the case $K<0$ in which the system is known to converge to a desynchronized configuration (see \cite{hong2011kuramoto}). 

For simplicity, in these simulations we only employed the GD-RBM algorithm, since using the GD approach we would obtain analogous results.

\begin{figure}[h!]
	\centering 
	\includegraphics[scale=1]{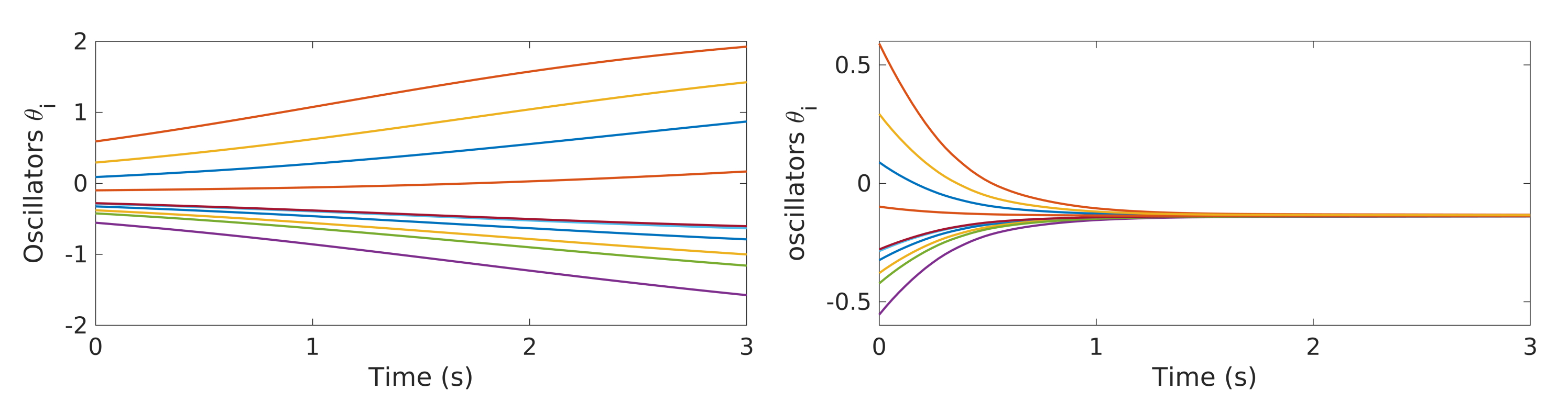}
	\caption{Evolution of the uncontrolled (left) and controlled (right) dynamics of the Kuramoto model \eqref{kuramoto_control} with $N=10$ oscillators and $K<0$.}\label{fig:7}
\end{figure}

We can see in Figure \ref{fig:7} how, in this case of a negative coupling gain, the uncontrolled dynamics is diverging as $t$ increases. On the other hand, when applying the control $\widehat{u}$, the system is once again steered to a synchronized configuration.

\begin{figure}[h!]
	\centering 
	\includegraphics[scale=0.4]{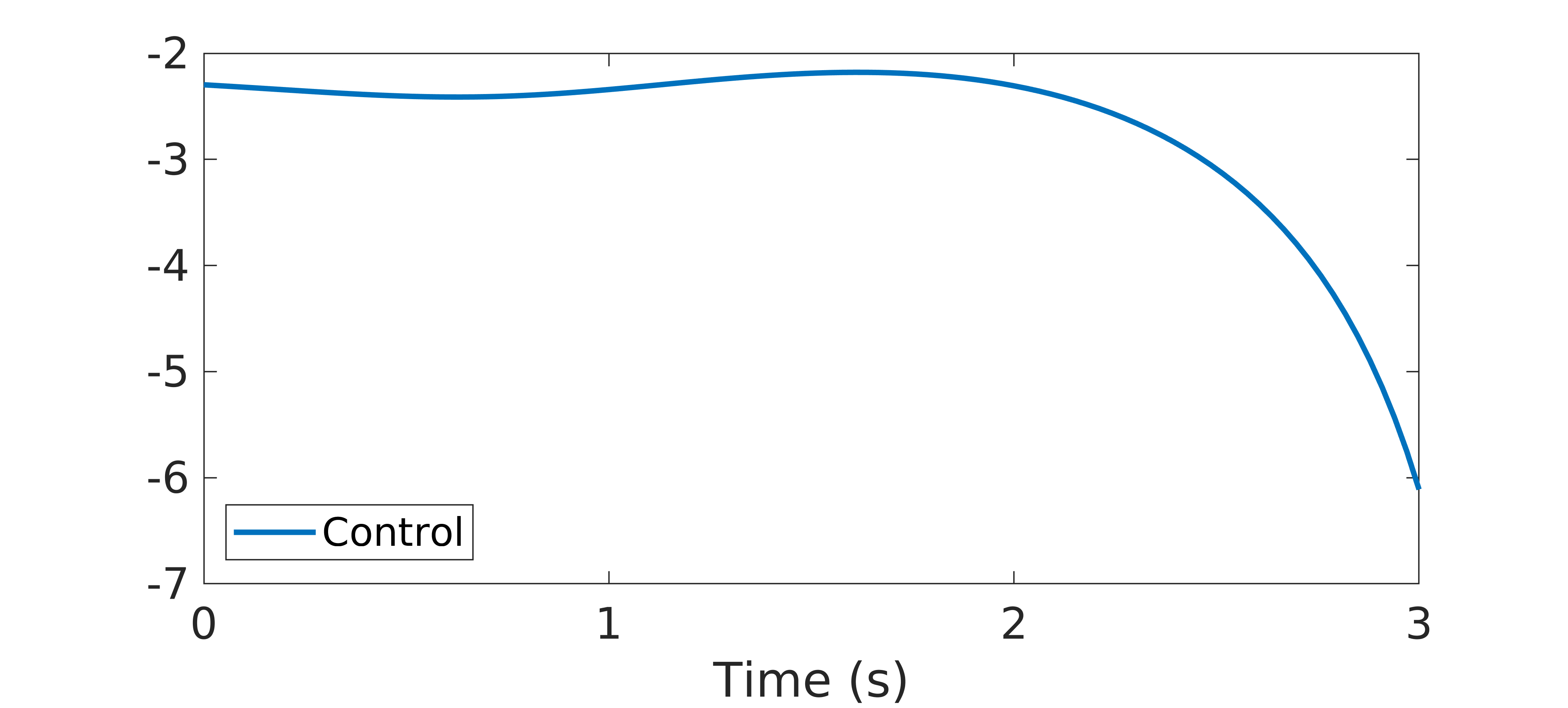}
	\caption{Control function $\widehat{u}$ obtained through the GD-RBM algorithm applied to the Kuramoto model \eqref{kuramoto_control} with $N=10$ oscillators and $K<0$.}\label{fig:8}
\end{figure}

At this regard, it is also interesting to observe that, this time, the control function we obtained is always negative (see Figure \ref{fig:8}). This fact is not surprising, if we recall that in \eqref{kuramoto_control} the control acts by modifying the coupling gain $K$ so that the oscillators are all synchronized at time $T$ and that, for the uncontrolled dynamics, synchronization requires $K>0$. 

Let us now complement our analysis by briefly showing the efficacy of our control strategy in a couple of more complex and realistic situations.

\subsubsection{The case of a sparse interaction network}\label{sparse_sub}

We start by considering the case of a sparse interaction network in our Kuramoto model \eqref{kuramoto_control}. In other words, we are considering here the following system
\begin{align}\label{kuramoto_control_sparse}
	\begin{cases}
		\displaystyle\dot{\theta}_i(t) = \omega_i + \frac{Ku(t)}{N}\sum_{j=1}^N a_{i,j}\sin\big(\theta_j(t)-\theta_i(t)\big),\quad i = 1,\ldots,N,\quad t>0
		\\
		\theta_i(0) = \theta^0_i,
	\end{cases}
\end{align}
with
\begin{align*}
	a_{i,j} = \begin{cases}
		1, & \mbox{ if } \theta_i \mbox{ is connected with } \theta_j
		\\
		0, & \mbox{ if } \theta_i \mbox{ is not connected with } \theta_j
	\end{cases}
\end{align*}

\begin{figure}[h!]
	\centering 
	\includegraphics[scale=0.4]{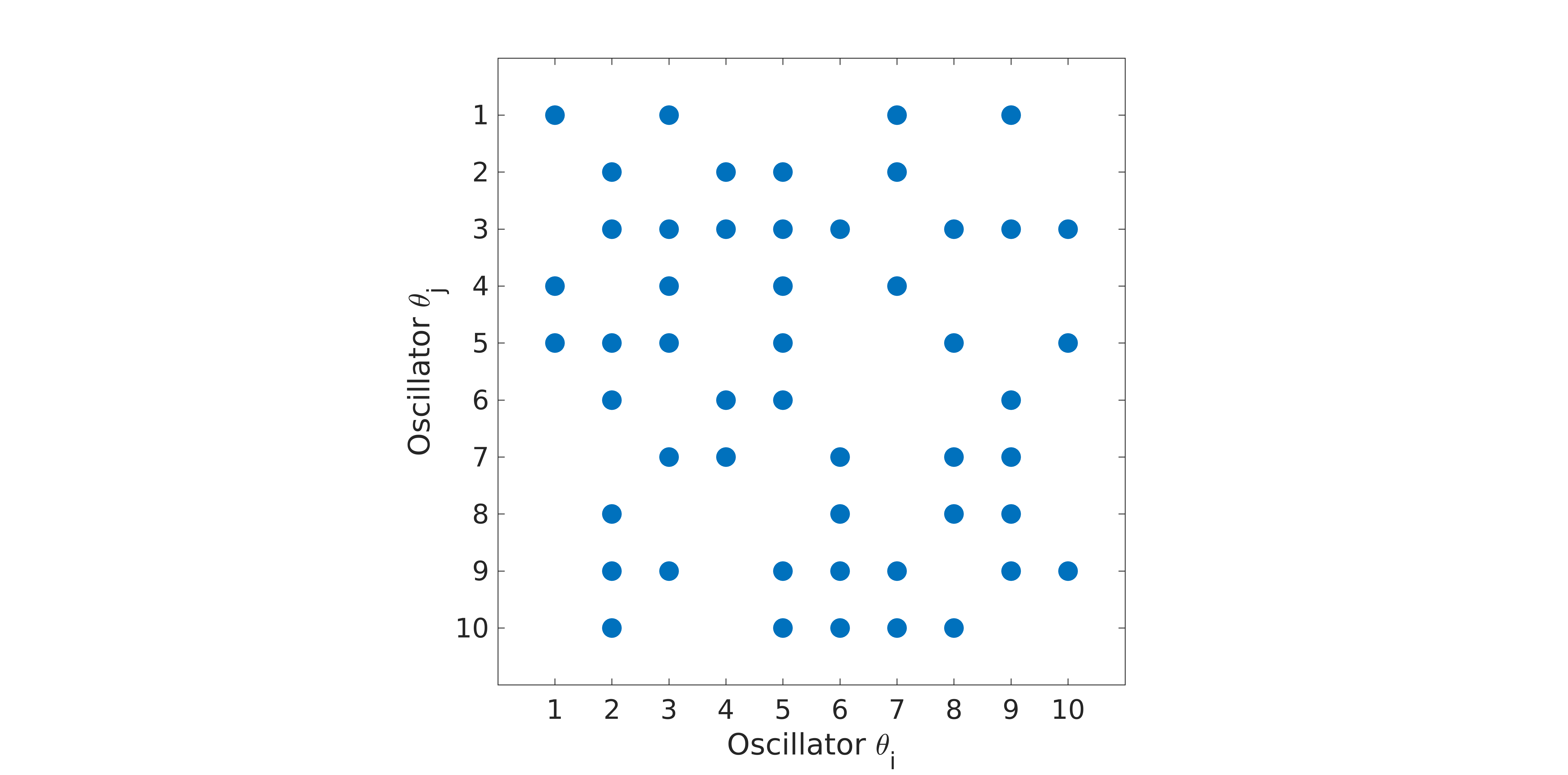}
	\caption{Sparse interaction scheme for the Kuramoto model \eqref{kuramoto_control_sparse}.}\label{fig:9}
\end{figure}

A schematic representation of the network considered in our simulations is given in Figure \ref{fig:9}, in which the blue dots correspond to $a_{i,j} = 1$.

The simulations have been performed with the same initial datum and time horizon we considered in our previous experiments. Moreover, we addressed here only the case of a strong coupling gain $K>K^\ast$. The minimization of the functional $J(u)$ has been performed with the GD algorithm.

\begin{figure}[h!]
	\centering 
	\includegraphics[scale=1]{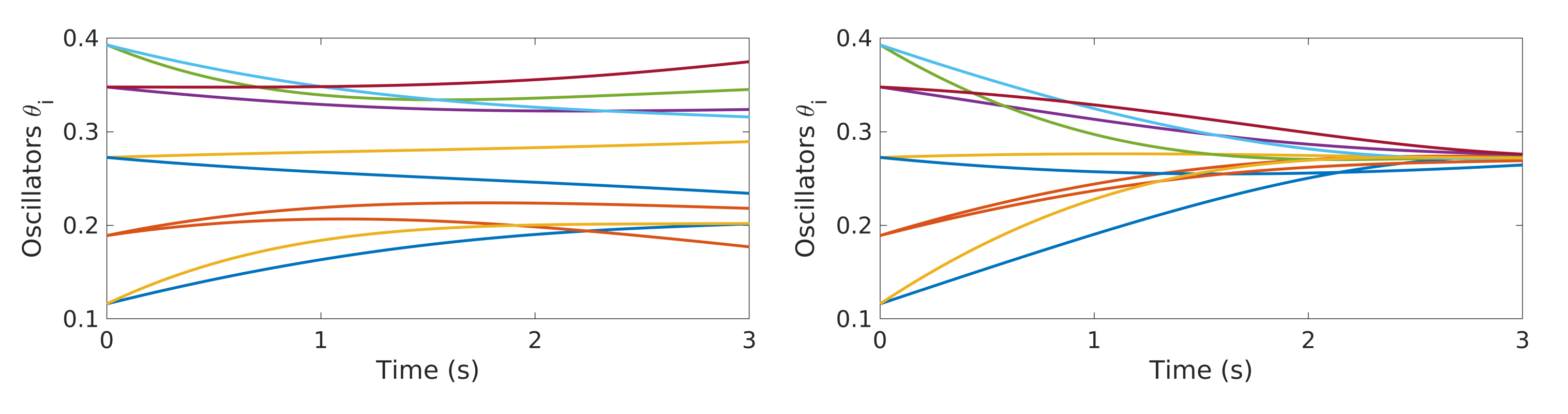}
	\caption{Evolution of the uncontrolled (left) and controlled (right) dynamics of the Kuramoto model \eqref{kuramoto_control_sparse} with $N=10$ oscillators, $K>K^\ast$ and interactions as in Figure \ref{fig:9}.}\label{fig:10}
\end{figure}

In Figure \ref{fig:10}, we show the evolution of the uncontrolled and controlled dynamics. As we can see, while in the absence of a control the oscillators are evolving towards a desynchronized configuration, when applying the control function we computed the system still reaches synchronization at time $T$.

\begin{figure}[h!]
	\centering 
	\includegraphics[scale=0.4]{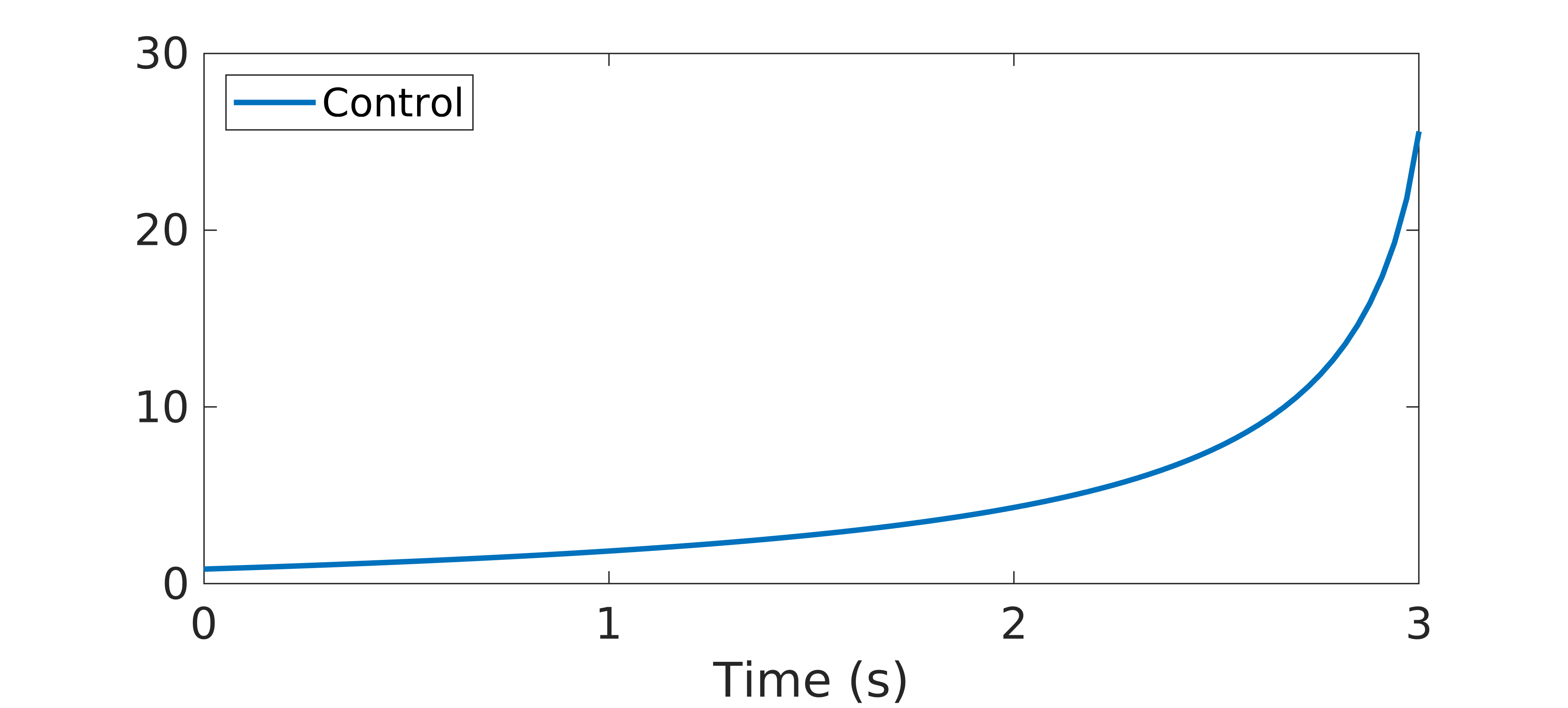}
	\caption{Control function $\widehat{u}$ for the Kuramoto model \eqref{kuramoto_control_sparse} with $N=10$ oscillators, $K>K^\ast$ and interactions as in Figure \ref{fig:9}.}\label{fig:11}
\end{figure}

The control function obtained for these numerical experiments is plotted in Figure \ref{fig:11}. We can observe how, differently from what is shown in Figures \ref{fig:4} and \ref{fig:5}, this time $\widehat{u}$ reaches larger values. This is not surprising, if we consider that now our model has a lower level of interactions and if we recall our previous discussion on how our control affects the Kuramoto dynamics.

\subsubsection{A second-order model with damping}\label{high_sub}

We consider here the second-order Kuramoto model with damping 
\begin{align*}
	\begin{cases}
		\displaystyle \ddot{\theta}_i(t) + \dot{\theta}_i(t) = \omega_i + \frac{Ku(t)}{N}\sum_{j=1}^N \sin\big(\theta_j(t)-\theta_i(t)\big), \quad i = 1,\ldots,N,\quad t>0
		\\
		\theta_i(0) = \theta^0_i, \quad \dot{\theta}_i(0) = \theta^1_i,
	\end{cases}
\end{align*}
which can be rewritten as the following first-order system 
\begin{align}\label{kuramoto_control_high}
	\begin{cases}
		\displaystyle \dot{\theta}_i(t) = \xi_i(t), & i = 1,\ldots,N,\quad t>0
		\\
		\displaystyle \dot{\xi}_i(t) = -\xi_i(t) + \omega_i + \frac{Ku(t)}{N}\sum_{j=1}^N \sin\big(\theta_j(t)-\theta_i(t)\big), & i = 1,\ldots,N,\quad t>0
		\\
		\theta_i(0) = \theta^0_i, \quad \xi_i(0) = \theta^1_i.
	\end{cases}
\end{align}

In the context of power grids, this model has been firstly introduced in the work \cite{filatrella2008analysis}. Later on, in \cite{schmietendorf2014self}, it has been extended to consider more complex scenarios in which the dynamics of the voltage amplitude is taken into account. 

Also in this case, we are interested in computing a control capable to steer the system to the synchronized configuration \eqref{consensus}. This can be done once again by solving the optimal control problem \eqref{functional}, this time under the dynamics \eqref{kuramoto_control_high}.

The simulations have been performed with the same initial datum $\Theta^0=(\theta_i^0)_{i=1}^N$ we considered in our previous experiments and with $\Theta^1=(\theta_i^1)_{i=1}^N = (0,0,\ldots,0)^\top$. The time horizon is once again $T=3s$. Moreover, we addressed here both the cases of a strong coupling gain $K>K^\ast$ and of a negative one $K<0$. The minimization of the functional $J(u)$ has been performed with the GD algorithm.

At this regard, we shall mention that, in \cite{tumash2019stability}, the GD methodology has been applied to obtain synchronization in a sparse network of Kuramoto oscillators with damping under the action of an additive control, i.e. the following model
\begin{align*}
	\begin{cases}
		\displaystyle \dot{\theta}_i(t) = \xi_i(t), & i = 1,\ldots,N,\quad t>0
		\\
		\displaystyle \dot{\xi}_i(t) = -\xi_i(t) + \omega_i + \frac{K}{N}\sum_{j=1}^N a_{i,j}\sin\big(\theta_j(t)-\theta_i(t)\big) + u_i(t), & i = 1,\ldots,N,\quad t>0
		\\
		\theta_i(0) = \theta^0_i, \quad \xi_i(0) = \theta^1_i.
	\end{cases}
\end{align*}

The advantages and disadvantages of this additive control action with respect to the multiplicative one we propose have been discussed in Section \ref{control_sec}. In particular, our control strategy allows us to deal also with negative coupling gain $K<0$, while in \cite{tumash2019stability} only $K>0$ has been considered.

\begin{figure}[h!]
	\centering
	\includegraphics[scale=1]{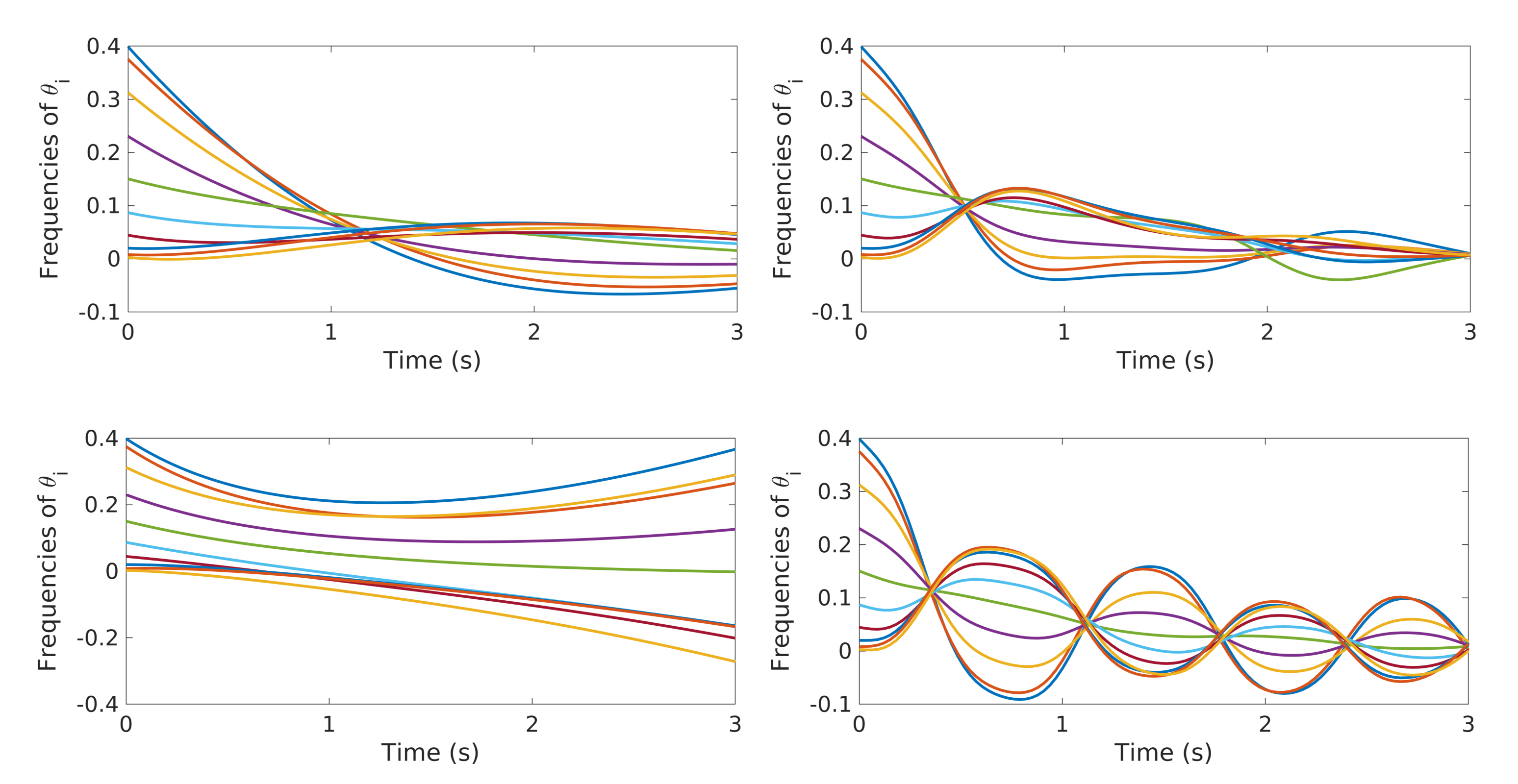}
	\caption{Evolution of the uncontrolled (left) and controlled (right) dynamics of the Kuramoto model \eqref{kuramoto_control_high} with $N=10$ oscillators and $K>K^\ast$ (top) and $K<0$ (bottom).}\label{fig:12}
\end{figure}

As a matter of fact, in Figure \ref{fig:12}, we show the evolution of the uncontrolled and controlled dynamics, for $K>K^\ast$ and $K<0$, respectively. Also in this case, the proposed control strategy allows us to compute an effective control function $\widehat{u}$ which steers the system to a synchronized configuration in time $T$.

\begin{figure}[h!]
	\centering
	\includegraphics[scale=1]{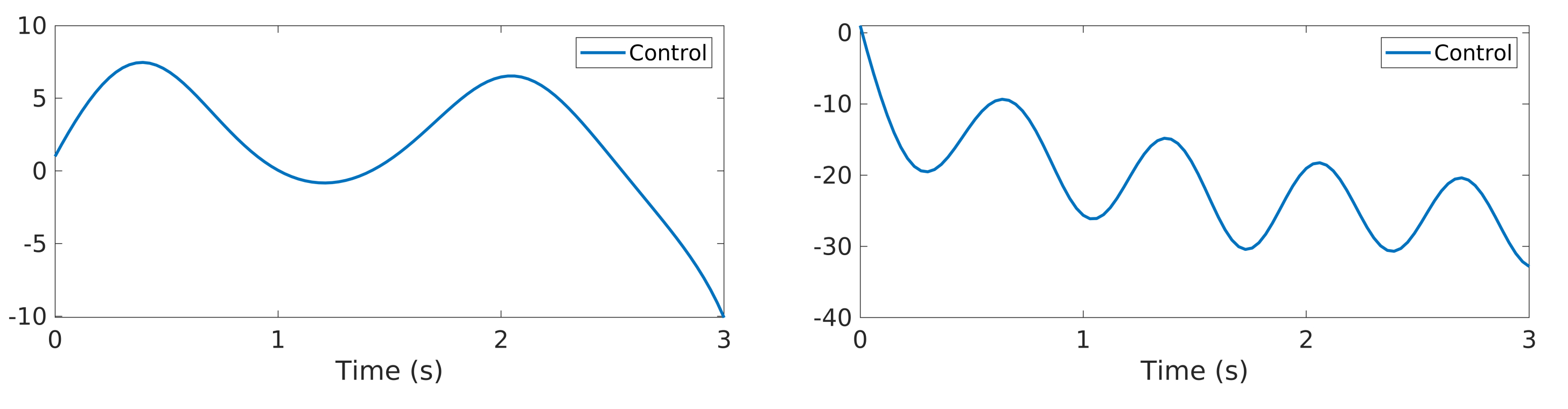}
	\caption{Control function $\widehat{u}$ for the Kuramoto model \eqref{kuramoto_control_high} with $N=10$ oscillators and $K>K^\ast$ (left) and $K<0$ (right).}\label{fig:14}
\end{figure}

Finally, the control functions obtained for these numerical experiments are plotted in Figure \ref{fig:14}. Also in these cases we can observe different behaviors than what is shown in Figures \ref{fig:4} and \ref{fig:5}. In particular, this time $\widehat{u}$ changes sign in the time horizon $(0,T)$. At this regard, let us mention that for the second-order Kuramoto model \eqref{kuramoto_control_high}, due to the presence of the damping term, our previous considerations on the sign of the control do not apply anymore. Hence, it is not surprising to obtain a behavior as the one displayed.

\subsection{Comparison of GD and GD-RBM}

We conclude this section on the numerical experiments by comparing the performances of the GD and GD-RBM algorithms for the computation of the optimal control $\widehat{u}$. 

To this end, we run simulations for increasing values of $N$, namely $N=10,50,100,250,1000$. As before, we chose a time horizon $T=3s$ and a penalization parameter $\beta = 10^{-7}$. Moreover, we considered the case of a large coupling gain $K>K^\ast$.

In what follows, we focus only on the simple case of the first-order Kuramoto model \eqref{kuramoto_control} with an all-to-all interaction network. We will briefly comment about possible extensions to the more realistic scenarios described in Sections \ref{sparse_sub} and \ref{high_sub} in the the last part of this paper, devoted to conclusions and open problems.

\begin{figure}[!h]
	\centering
	\includegraphics[scale=0.3]{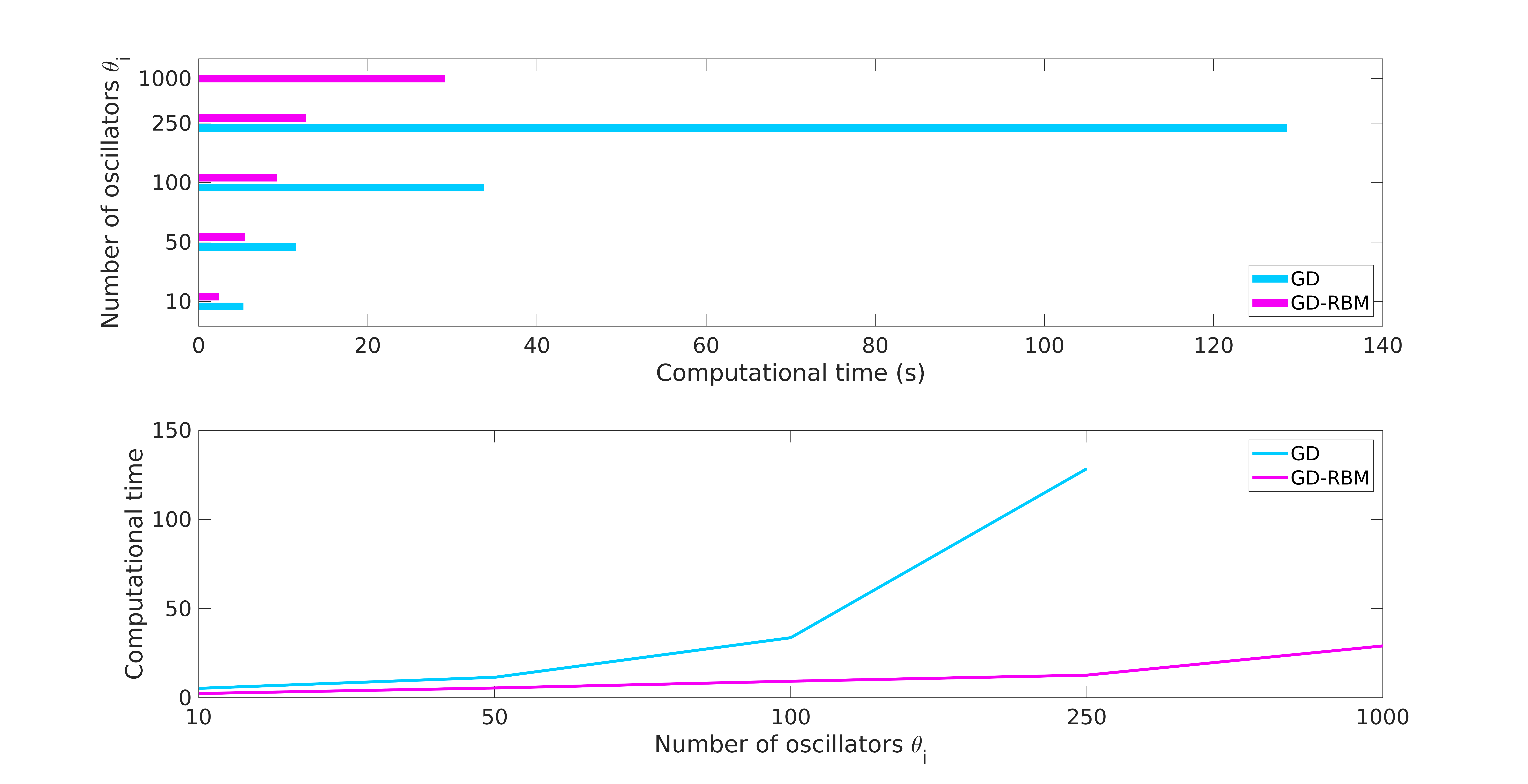}
	\caption{Computational times required by the GD and GD-RBM algorithm to compute the optimal control $\widehat{u}$ with increasing values of $N$.}\label{fig:15} 
\end{figure}

In Table \ref{timeTable1} (see also Figure \ref{fig:15}) we collected the computational times required by the two methodologies to solve the optimization problem \eqref{functional}. At this regard, let us stress that the values contained in the table do not represent the time required by the control to synchronize the network which, we recall, is a fixed external input in our algorithms. 

\begin{center} 
	\begin{table}[h]
		\renewcommand{\arraystretch}{1.2}
		\begin{tabular}{|c|c|c|}
			\hline & GD & GD-RBM
			\\
			\hline $N$ & Time (sec.) & Time (sec.) 
			\\
			\hline $10$ & $5.3$ & $2.4$ 
			\\
			\hline $50$ & $11.5$ & $5.5$
			\\
			\hline $100$ & $33.7$ & $9.3$
			\\
			\hline $250$ & $128.5$ & $12.7$ 
			\\
			\hline $1000$ & & $29.1$ 
			\\
			\hline
		\end{tabular}\caption{Computational times required by the GD and GD-RBM algorithm to compute the optimal control $\widehat{u}$ with increasing values of $N$.}\label{timeTable1}
	\end{table}
\end{center} 

Our simulations show how, for low values of $N$, the two approaches show similar behaviors. Nevertheless, when increasing the number of oscillators in our system, the advantages of the GD-RBM methodology with respect to GD become evident. In particular, the growth of the computational time for GD-RBM is significantly less pronounced than for GD. As a matter of fact, in the case of $N=1000$ oscillators, we decided not to perform the simulations with the GD algorithm since its behavior with smaller values of $N$ already suggested that this experiment would be computationally too expensive.

\begin{figure}[!h]
	\centering 
	\includegraphics[scale=0.45]{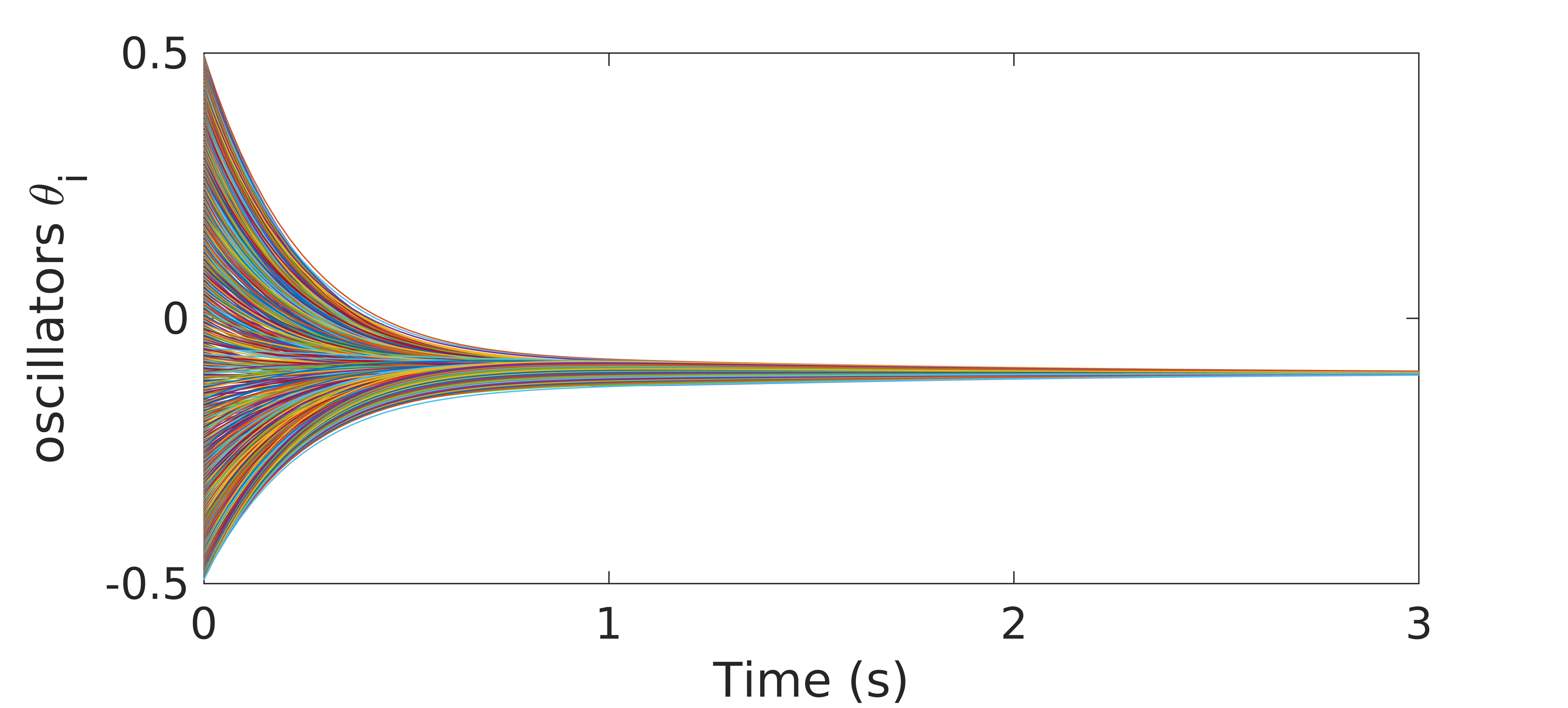}
	\caption{Evolution of the controlled dynamics of the Kuramoto model \eqref{kuramoto_control} with $N=1000$ oscillators. The control has been computed with the GD-RBM algorithm.}\label{fig:16}
\end{figure}

On the other hand, even with $N=1000$ oscillators in the system, the GD-RBM approach turns out to be able to compute an effective control for the Kuramoto model \eqref{kuramoto_control} (see Figure \ref{fig:16}) in about $29$ seconds. 

\section{Conclusions}\label{conclusions_sec}

This paper deals with the synchronization of coupled oscillators described by the Kuramoto model. In particular, we design a unique scalar control function $u(t)$ capable of steering a $N$-dimensional network of oscillators to a synchronized configuration in a finite time horizon. This is done following a standard optimal control approach and obtaining the function $u(t)$ via the minimization of a suitable cost functional. 

With this approach, we computed a control which acts as a multiplicative force enhancing the coupling between the oscillators in the network, thus favoring the synchronization in the time horizon provided. 

To carry out this minimization process, we used a Gradient Descent (GD) methodology, commonly employed in the optimal control community, which we have coupled with the novel Random Batch Method (RBM) for a more efficient numerical resolution of the Kuramoto dynamics. The main purpose of this work has been to show how the introduction of RBM into GD may yield considerable improvement in term of the computational complexity, in particular for large oscillator networks.

Our simulations results have shown the following main facts:
\begin{itemize}
	\item The proposed control strategy is indeed effective to reach a synchronized configuration in a finite time horizon. Moreover, it allows to deal efficiently with large networks of oscillators (namely, $N=1000$) and with the case of a low coupling gain in the network, when the uncontrolled dynamics is not expected to reach a synchronized configuration.
	\item For large values of $N$, the inclusion of RBM into the GD algorithm significantly reduces the computational burden for the control computation, thus allowing to deal with high-dimensional oscillators networks in a more efficient way. 
\end{itemize}

In conclusion, the study conducted in this paper suggests that the proposed methodology, based on the combination of the standard GD optimization algorithm with the novel RBM method for the numerical resolution of multi-agent dynamics, may help in significantly reduce the computational complexity for the control computation of the Kuramoto model, in particular in the case of a high-dimensional system. 

At this regard, we shall stress that the analysis in this paper has been developed mostly in a simplified framework of a network with all-to-all coupling, in which the oscillators are all interacting among themselves. The following interesting questions remain unaddressed:
\begin{itemize}
	\item To study whether our methodology remains valid in more complex scenarios of networks with sparse interaction topologies, perturbations due to disconnections, or rewiring. In particular, it would be relevant to determine if the reduction in computational complexity that we obtained through the GD-RBM algorithm is related with the density of the network or, instead, this approach allows to deal efficiently also with the case of a low number of interactions. A starting point for this analysis would be to determine whether the GD-RBM methodology can be successfully applied in the scenarios we addressed in Sections \ref{sparse_sub} and \ref{high_sub}.
	\item To analyze what happens if, instead of selecting the oscillators uniformly at random during the batching process in RBM, we organize them in groups with similar frequencies. At this regard, it would be important to understand if the methodology we propose is still effective or, instead, some modifications need to be introduced.	
	\item To analyze if our methodology may be applied also in different framework than the ones considered in this paper. For instance, a similar formalism than the one we have proposed has been developed for computing rare-events in oscillator networks due to noise. See for instance \cite{hindes2018rare,hindes2019network}. In these mentioned contributions, the objective functional is the probability for a rare event to happen. The actions are more complicated than the simple $L^2$-norm, but the batch techniques we employed may be useful in this context as well.	
\end{itemize}

All these are key open problem which will be considered in future works. 

\section*{Acknowledgment}

The authors wish to acknowledge Jes\'us Oroya and Dongnam Ko (Chair of Computational Mathematics, Fundaci\'on Deusto, Bilbao, Spain) for interesting discussion on the topics of the present paper.

\bibliographystyle{acm}
\bibliography{paper_Kuramoto_arxiv}

\begin{thebibliography}{10}

\bibitem{acebron2005kuramoto}
{\sc Acebr{\'o}n, J.~A., Bonilla, L.~L., Vicente, C. J.~P., Ritort, F., and
  Spigler, R.}
\newblock The {K}uramoto model: A simple paradigm for synchronization
  phenomena.
\newblock {\em Rev. Modern Phys. 77}, 1 (2005), 137.

\bibitem{ben2005opinion}
{\sc Ben-Naim, E.}
\newblock Opinion dynamics: rise and fall of political parties.
\newblock {\em Europhys. Lett. 69}, 5 (2005), 671.

\bibitem{biccari2019dynamics}
{\sc Biccari, U., Ko, D., and Zuazua, E.}
\newblock Dynamics and control for multi-agent networked systems: A
  finite-difference approach.
\newblock {\em Math. Models Methods Appl. Sci. 29}, 04 (2019), 755--790.

\bibitem{bottou2018optimization}
{\sc Bottou, L., Curtis, F., and Nocedal, J.}
\newblock Optimization methods for large-scale machine learning.
\newblock {\em SIAM Rev. 60}, 2 (2018).

\bibitem{buck1988synchronous}
{\sc Buck, J.}
\newblock Synchronous rhythmic flashing of fireflies {II}.
\newblock {\em Quart. Rev. Bio. 63}, 3 (1988), 265--289.

\bibitem{chassin2005evaluating}
{\sc Chassin, D.~P., and Posse, C.}
\newblock Evaluating north american electric grid reliability using the
  {B}arab{\'a}si-{A}lbert network model.
\newblock {\em Phys. A 355}, 2-4 (2005), 667--677.

\bibitem{chopra2005synchronization}
{\sc Chopra, N., and Spong, M.~W.}
\newblock On synchronization of {K}uramoto oscillators.
\newblock In {\em Proceedings of the 44th IEEE Conference on Decision and
  Control\/} (2005), IEEE, pp.~3916--3922.

\bibitem{chopra2006passivity}
{\sc Chopra, N., and Spong, M.~W.}
\newblock Passivity-based control of multi-agent systems.
\newblock In {\em Adv. Robot Control}. Springer, 2006, pp.~107--134.

\bibitem{chopra2009exponential}
{\sc Chopra, N., and Spong, M.~W.}
\newblock On exponential synchronization of {K}uramoto oscillators.
\newblock {\em IEEE Trans. Autom. Control 54}, 2 (2009), 353--357.

\bibitem{dong2013synchronization}
{\sc Dong, J.-G., and Xue, X.}
\newblock Synchronization analysis of {K}uramoto oscillators.
\newblock {\em Commun. Math. Sci. 11}, 2 (2013), 465--480.

\bibitem{dorfler2010synchronization}
{\sc D{\"o}rfler, F., and Bullo, F.}
\newblock Synchronization of power networks: Network reduction and effective
  resistance.
\newblock {\em IFAC Proceedings Volumes 43}, 19 (2010), 197--202.

\bibitem{dorfler2013synchronization}
{\sc D{\"o}rfler, F., Chertkov, M., and Bullo, F.}
\newblock Synchronization in complex oscillator networks and smart grids.
\newblock {\em Proc. Nat. Acad. Sci. 110}, 6 (2013), 2005--2010.

\bibitem{filatrella2008analysis}
{\sc Filatrella, G., Nielsen, A.~H., and Pedersen, N.~F.}
\newblock Analysis of a power grid using a {K}uramoto-like model.
\newblock {\em Europ. Phys. J. B 61}, 4 (2008), 485--491.

\bibitem{ha2015remarks}
{\sc Ha, S.-Y., Kim, H.~K., and Park, J.}
\newblock Remarks on the complete synchronization of {K}uramoto oscillators.
\newblock {\em Nonlinearity 28}, 5 (2015), 1441.

\bibitem{ha2016emergence}
{\sc Ha, S.-Y., Kim, H.~K., and Ryoo, S.~W.}
\newblock Emergence of phase-locked states for the {K}uramoto model in a large
  coupling regime.
\newblock {\em Commun. Math. Sci. 14}, 4 (2016), 1073--1091.

\bibitem{hindes2019network}
{\sc Hindes, J., Jacquod, P., and Schwartz, I.~B.}
\newblock Network desynchronization by non-gaussian fluctuations.
\newblock {\em Phys. Rev. E 100}, 5 (2019), 052314.

\bibitem{hindes2018rare}
{\sc Hindes, J., and Schwartz, I.~B.}
\newblock Rare slips in fluctuating synchronized oscillator networks.
\newblock {\em Chaos 28}, 7 (2018), 071106.

\bibitem{hong2011kuramoto}
{\sc Hong, H., and Strogatz, S.~H.}
\newblock Kuramoto model of coupled oscillators with positive and negative
  coupling parameters: an example of conformist and contrarian oscillators.
\newblock {\em Phys. Rev. Lett. 106}, 5 (2011), 054102.

\bibitem{jadbabaie2004stability}
{\sc Jadbabaie, A., Motee, N., and Barahona, M.}
\newblock On the stability of the {K}uramoto model of coupled nonlinear
  oscillators.
\newblock In {\em Proc. 2004 American Control Conference\/} (2004), vol.~5,
  IEEE, pp.~4296--4301.

\bibitem{jin2020random}
{\sc Jin, S., Li, L., and Liu, J.-G.}
\newblock Random {B}atch {M}ethods ({RBM}) for interacting particle systems.
\newblock {\em J. Comput. Phys. 400\/} (2020), 108877.

\bibitem{kuramoto1975self}
{\sc Kuramoto, Y.}
\newblock Self-entrainment of a population of coupled non-linear oscillators.
\newblock In {\em International symposium on mathematical problems in
  theoretical physics\/} (1975), Springer, pp.~420--422.

\bibitem{markdahl2020high}
{\sc Markdahl, J., Thunberg, J., and Goncalves, J.}
\newblock High-dimensional {K}uramoto models on {S}tiefel manifolds synchronize
  complex networks almost globally.
\newblock {\em Automatica 113\/} (2020), 108736.

\bibitem{mehyar2005distributed}
{\sc Mehyar, M., Spanos, D., Pongsajapan, J., Low, S.~H., and Murray, R.~M.}
\newblock Distributed averaging on asynchronous communication networks.
\newblock In {\em Proceedings of the 44th IEEE Conference on Decision and
  Control\/} (2005), IEEE, pp.~7446--7451.

\bibitem{nabi2011single}
{\sc Nabi, A., and Moehlis, J.}
\newblock Single input optimal control for globally coupled neuron networks.
\newblock {\em J. Neural Eng. 8}, 6 (2011), 065008.

\bibitem{nesterov2004applied}
{\sc Nesterov, Y.}
\newblock {\em Applied optimization}.
\newblock Introductory Lectures on Convex Optimization: A Basic Course. Kluwer
  Academic Publishers Boston, Dordrecht, London, 2004.

\bibitem{nocedal2006numerical}
{\sc Nocedal, J., and Wright, S.}
\newblock {\em Numerical optimization}.
\newblock Springer Science \& Business Media, 2006.

\bibitem{olfati2006flocking}
{\sc Olfati-Saber, R.}
\newblock Flocking for multi-agent dynamic systems: Algorithms and theory.
\newblock {\em IEEE Trans. Autom. Control 51}, 3 (2006), 401--420.

\bibitem{olfati2007consensus}
{\sc Olfati-Saber, R., Fax, J.~A., and Murray, R.~M.}
\newblock Consensus and cooperation in networked multi-agent systems.
\newblock {\em Proc. IEEE 95}, 1 (2007), 215--233.

\bibitem{rosenblum2004delayed}
{\sc Rosenblum, M., and Pikovsky, A.}
\newblock Delayed feedback control of collective synchrony: An approach to
  suppression of pathological brain rhythms.
\newblock {\em Phys. Rev. E 70}, 4 (2004), 041904.

\bibitem{sachtjen2000disturbances}
{\sc Sachtjen, M., Carreras, B., and Lynch, V.}
\newblock Disturbances in a power transmission system.
\newblock {\em Phys. Rev. E 61}, 5 (2000), 4877.

\bibitem{schmidhuber2015deep}
{\sc Schmidhuber, J.}
\newblock Deep learning in neural networks: an overview.
\newblock {\em Neural networks 61\/} (2015), 85--117.

\bibitem{schmietendorf2014self}
{\sc Schmietendorf, K., Peinke, J., Friedrich, R., and Kamps, O.}
\newblock Self-organized synchronization and voltage stability in networks of
  synchronous machines.
\newblock {\em Eur. Phys. J. Spec. Top. 223}, 12 (2014), 2577--2592.

\bibitem{sepulchre2005collective}
{\sc Sepulchre, R., Paley, D., and Leonard, N.}
\newblock Collective motion and oscillator synchronization.
\newblock In {\em Cooperative control}. Springer, 2005, pp.~189--205.

\bibitem{shalev2014accelerated}
{\sc Shalev-Shwartz, S., and Zhang, T.}
\newblock Accelerated proximal stochastic dual coordinate ascent for
  regularized loss minimization.
\newblock In {\em International Conference on Machine Learning\/} (2014),
  pp.~64--72.

\bibitem{strogatz2001exploring}
{\sc Strogatz, S.~H.}
\newblock Exploring complex networks.
\newblock {\em Nature 410}, 6825 (2001), 268--276.

\bibitem{strogatz2005crowd}
{\sc Strogatz, S.~H., Abrams, D.~M., McRobie, A., Eckhardt, B., and Ott, E.}
\newblock Crowd synchrony on the millennium bridge.
\newblock {\em Nature 438}, 7064 (2005), 43--44.

\bibitem{sun2009master}
{\sc Sun, J., Bollt, E.~M., and Nishikawa, T.}
\newblock Master stability functions for coupled nearly identical dynamical
  systems.
\newblock {\em Europhys. Lett. 85}, 6 (2009), 60011.

\bibitem{taylor2016synchronization}
{\sc Taylor, D., Skardal, P.~S., and Sun, J.}
\newblock Synchronization of heterogeneous oscillators under network
  modifications: Perturbation and optimization of the synchrony alignment
  function.
\newblock {\em SIAM J. Appl. Math. 76}, 5 (2016), 1984--2008.

\bibitem{toscher2010collaborative}
{\sc Toscher, A., and Jahrer, M.}
\newblock Collaborative filtering applied to educational data mining.
\newblock {\em KDD cup\/} (2010).

\bibitem{trelat2005controle}
{\sc Tr{\'e}lat, E.}
\newblock {\em Contr{\^o}le optimal: th{\'e}orie \& applications}.
\newblock Vuibert, Collection Math{\'e}matiques Concr{\`e}tes, 2005.

\bibitem{troltzsch2010optimal}
{\sc Tr{\"o}ltzsch, F.}
\newblock {\em Optimal control of partial differential equations: theory,
  methods, and applications}.
\newblock American Mathematical Soc., 2010.

\bibitem{tukhlina2007feedback}
{\sc Tukhlina, N., Rosenblum, M., Pikovsky, A., and Kurths, J.}
\newblock Feedback suppression of neural synchrony by vanishing stimulation.
\newblock {\em Phys. Rev. E 75}, 1 (2007), 011918.

\bibitem{tumash2019stability}
{\sc Tumash, L., Olmi, S., and Sch{\"o}ll, E.}
\newblock Stability and control of power grids with diluted network topology.
\newblock {\em Chaos: An Interdisciplinary Journal of Nonlinear Science 29}, 12
  (2019), 123105.

\bibitem{walker1969acoustic}
{\sc Walker, T.~J.}
\newblock Acoustic synchrony: two mechanisms in the snowy tree cricket.
\newblock {\em Science 166}, 3907 (1969), 891--894.

\bibitem{wiesenfeld1998frequency}
{\sc Wiesenfeld, K., Colet, P., and Strogatz, S.~H.}
\newblock Frequency locking in {J}osephson arrays: connection with the
  {K}uramoto model.
\newblock {\em Phys. Rev. E 57}, 2 (1998), 1563.

\bibitem{winfree1967biological}
{\sc Winfree, A.~T.}
\newblock Biological rhythms and the behavior of populations of coupled
  oscillators.
\newblock {\em J. Theor. Bio. 16}, 1 (1967), 15--42.

\end{thebibliography}

\end{document}